\documentclass[10pt,superscriptaddress, twocolumn,showkeys,aps,prb]{revtex4-1}
\usepackage{cancel}
\usepackage{mathrsfs}
\usepackage{bbold}
\usepackage{float}
\usepackage{graphicx}
\usepackage{bm}
\usepackage{amsmath} 
\usepackage{booktabs}
\usepackage{color}

\def\<{\langle}
\def\>{\rangle}
\begin{document}
\title{Quantum oscillation and wave packet revival in conical graphene structure}
\author{Debabrata Sinha}
\affiliation{TIFR Centre for Interdisciplinary Sciences, Hyderabad 500075, India}
\author{Bertrand Berche}
\affiliation{Statistical Physics Group, Institut Jean Lamour, UMR CNRS No 7198, Universit\'e de Lorraine, B.P. 70239, F-54506 Vand\oe uvre les Nancy, France }
\email{debabratas@tifrh.res.in}
\email{bertrand.berche@univ-lorraine.fr}

\begin{abstract}
We present analytical expressions for the eigenstates and eigenvalues of electrons confined in a graphene monolayer in the presence of a disclination. The calculations are performed in the continuum limit approximation in the vicinity of the Dirac points, solving Dirac equation by freezing out the carrier radial motion. We include the effect of an external magnetic field and show the appearence of Aharonov-Bohm oscillation and find out the conditions of gapped and gapless states in the spectrum. We show that the gauge field due to a disclination lifts the orbital degeneracy originating from the existence of two valleys. The broken valley degeneracy has a clear signature on quantum oscillations and wave packet dynamics.
\end{abstract}

\maketitle
\section{Introduction}
The presence of topological defects in condensed matter systems can modify thoroughly their physical properties~\cite{Guinea1}.
From the point of view of the geometrical effects, topological defects can induce curvature and/or torsion in the geometrical background. Differential geometry can thus be used 
for the study of a variety of physical systems, ranging from superfluid helium and its numerous spontaneous symmetry breakings~\cite{Volovik}, to 
liquid crystals~\cite{Fumeron} or mesoscopic physics~\cite{Guinea1,Sinha}. The prominent work of Katanaev and Volovich~\cite{Katanaev} has showed the efficiency of this geometric approach to
investigate the propagation of elastic waves in the presence of topological defects such as screw dislocations, edge dislocations or disclinations. 

In a recent paper, one of us showed how the presence of a screw dislocation can generate spin current in an electron gas~\cite{Sinha}. A screw dislocation  is a source of torsion. Torsion can also be built in directly in the physical structure, like it is the case in chiral molecules where chirality couples to spin in the presence of spin-orbit interaction and can be used for spin selectivity applications,  e.g. to polarize an electron beam~\cite{Medina1} or to exalt a spin orientation of a spin current propagating through the molecule~\cite{Medina2}.  Edge dislocations are also the source of torsion, and their role
was studied in topological insulators~\cite{Schmeltzer}.

It is desirable to consider in this context  topological defects which generate curvature.   Single disclinations are line sources of curvature and they correspond to the generation of a conical geometry via the Volterra construction.
Our aim in the present paper is to investigate the properties of a graphene layer in which the crystal symmetry is locally modified (a hexagon is replaced by a pentagon or a square for example), giving rise to large scale properties analogous to the conical geometry mentioned above. 
The desire to employ graphene for studying curvature effects is motivated by the simplicity of the Hamiltonian and the important potential applications of graphene in naoelectronics and possibly future quantum computing devices \cite{Castro}. Carbon nanotubes are other famous elementary exemples of curvature effects in a carbon sheets, but the case of disclinations is different due to the presence of a singularity which sits at the defect location.

Recently, there {have} been many studies on quantum rings, in which the confinement of electrons with phase coherence of wave function gives rise the Aharonov-Bohm \cite{Aharonov_Bohm} and Aharonov-Casher effects \cite{Aharonov_Casher,MLB08}. Quantum rings have been investigated both {experimentally and theoretically} in semiconductor devices~\cite{Lipparini}
 and also in graphene layers both in monolayers and bilayers \cite{RecherEtAl,Farias,Farias_Erratum,Romera,BMB}. In this context it is important to study the curvature effects on electron properties e.g. charge currents or quantum oscillations in a graphene monolayer ring.

Graphene also offers a way to probe  quantum field theory in condensed matter systems, as its low energy spectrum is described by the Dirac Hamiltonian of massless fermions. The Dirac electrons in graphene occur in two degenerate families which correspond to the presence of two different valleys in the band structure -- a phenomenon known as {\em fermion doubling}. Due to the valley degeneracy, in many cases, it is difficult to observe the intrinsic physics of a single valley in experiments, because the two valleys have equal and opposite contributions to a measurable quantity. One way to break the valley degeneracy is the production of fictitious magnetic field in a single valley by a lattice defect. The field has an opposite sign in the other valley, so it lifts the degeneracy. The broken valley degeneracies leads to some interesting features on quantum oscillations and revival times of wave packets, which are the main concern of this paper. We show that the signature of the broken valley degeneracy is clearly visible in the persistent current and the wave packet characteristic oscillation times.

{Here, we consider a model of graphene with a wedge-dislocation, which can be understood from Volterra's cut-and-glue constructions. Such conical defects are elementary objects which are observed in graphene layers and in various carbon based structures\cite{Gonzalez_Guinea,Lammert_Crespi,Cortijo_Vozmediano,Yazyev_Louie,Cockayne_Rutter}. Our calculations are based on the continuum model where fictitious gauge fields due to the defects are coupled to the Dirac fermions~\cite{Guinea1,Ruegg}. We solve this model by freezing out the carriers'radial motion, keeping into account only their angular motion in a effectively one-dimensional system (namely a quantum ring).} The full calculations including carriers radial motion will be presented elsewhere \cite{sinha_berche}.

{The paper is organized as follows: in Sec.~\ref{Theory} we present the theoretical model and numerical results for the energy spectrum, pseudo-spin polarization and charge current. In Sec.~\ref{Wave Packet Revival}, the revivals and typical time scales of electrons wave function are presented. Sec.~\ref{Summary} contains a summary of the main results and conclusions.}

\section{Theory}\label{Theory}
\begin{figure}[ht]
\rotatebox{0}{\includegraphics[width=3.2in]{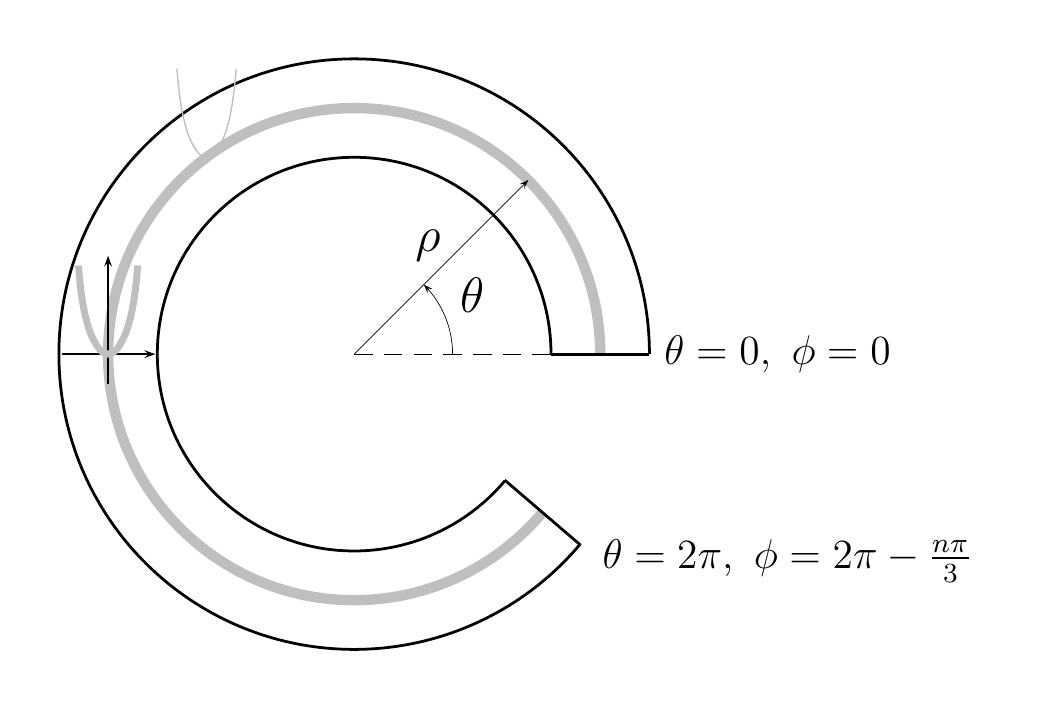}}
\caption{Unfolded plane of lattice where a wedge of angle $n\pi/3$ is removed ($n=1$ here). A potential (sketch on the left part of the ring) confines the electrons on a ring of radius $\rho$, avoiding the singularity at the origin. We rescaled the angle $\phi$ of the unfolded plane to the new angle $\theta=\phi/(1-\frac{n}{6})$.}
\label{wedge-disclination}
\end{figure}

In the low energy limit, the Hamiltonian of a graphene layer is given by
\begin{eqnarray}
H=v_{F}[\tau_{z}\sigma_{x}p_{x}+\sigma_{y}p_{y}]+\Delta\tau_{z}\sigma_{z}
\label{Low-Hamil}
\end{eqnarray}
where $\vec{\sigma}=(\sigma_{x},\sigma_{y},\sigma_{z})$, $\vec\tau=(\tau_{x},\tau_{y},\tau_{z})$ are Pauli matrices denoting the sublattice and valley degrees of freedom, respectively, and $v_{F}$ is the Fermi velocity. The finite mass term $\Delta$, which might be caused by an interaction with the underlying substrate or by spin-orbit coupling is also taken into account\cite{McCann,Recher}. The Hamiltonian (\ref{Low-Hamil}) acts on the four-component spinors
\begin{eqnarray}
\Psi(\vec r)=
(\psi_{A+}(\vec r)\ \psi_{B+}(\vec r)\ \psi_{A-}(\vec r)\ \psi_{B-}(\vec r) )^T.
\end{eqnarray} 
The pseudospin indices $A$ and $B$ label the two sublattices of graphene and the valley indices $+$ and $-$ refer to the two inequivalent Dirac point $K$ and $K'$ in the Brillouin zone. In a two dimensional plane for a flat graphene, the angular boundary condition for a Dirac spinor as it goes a closed path is given by
\begin{eqnarray}
\Psi(r,\phi=2\pi)=e^{i\pi\sigma_{z}\tau_{z}}\Psi(r,\phi=0)
\label{bc}
\end{eqnarray}
The presence of defects like disclination modifies the angular boundary conditions in Eq.(\ref{bc}). It is indeed known that deformation in the honeycomb lattice enter  in the continuum limit description  as fictitious gauge fields coupled to the electron momentum as a vector potential  \cite{Guinea1,Guinea2,Viswanath}. We introduced a wedge disclination via the Volterra construction: first a sector of $n\pi/3$ is removed or added on the flat graphene lattice and this is then followed by the gluing the edges of the cut (see Fig.~\ref{wedge-disclination}). An index $n$ defines the type of disclination in the honeycomb lattice. Keeping the symmetry of the lattice, a hexagon can now be replaced by pentagon ($n=1$), square ($n=2$), heptagon ($n=-1$) and octagon ($n=-2$) by the cut-and-glue procedure.

{From Eq.(\ref{bc}), the wave function is multiplied by the factor $e^{i\phi\sigma_{z}\tau_{z}/2}$ to keep the Hamiltonian covariance. So, when a sector is removed from the graphene plane, the wave function should transform according to}
\begin{eqnarray}
\Psi(r,\theta=2\pi)=-e^{i2\pi[1-(n/6)]\sigma_{z}\tau_{z}/2}\Psi(r,\theta=0)
\end{eqnarray} 
{When $n$ is odd, an additional phase is required for the parallel transport of a Dirac fermion since when the spinor is transported around the apex, 
the phase mismatch due to the sublattice permutation (as for example a $B$ site comes to a $B$ site instead of $A$) must be compensated to preserve the complete wave function single-valued. Considering all the factors, the angular boundary condition for a general $n$ now becomes \cite{RueggSM,Lammert_Crespi}}
\begin{eqnarray}
\Psi(r,\theta=2\pi)=-e^{i2\pi[-(n\sigma_{y}\tau_{y}/4)+(1-(n/6))\sigma_{z}\tau_{z}/2]}\Psi(r,\theta=0)
\label{mod_bc}
\end{eqnarray}
Note that for even $n$, the matrix $\tau_{y}$ does not play any role. Eq.(\ref{mod_bc}) can be written as
\begin{eqnarray}
\Psi(r,\theta=2\pi)={W}_{n}\Psi(r,\theta=0)
\label{mod-bc}
\end{eqnarray}
where 
\begin{equation}{W}_{n}=e^{i\frac{n\pi}{3}(\sigma_{z}\tau_{z}-3\sigma_{y}\tau_{y}/2)}.\end{equation} 
Where we rescale the angle $\phi$ of the unfolded plane to the new angle $\theta=\phi/(1-\frac{n}{6})$, where $\theta$ is varying from $0$ to $2\pi$. The general gauge transformation given in  Eq.~(\ref{mod-bc}) for any type of disclination is defined by the index $n=\pm 1, \pm 2$ defined above. 
For a local gauge transformation, Eq.~(\ref{mod-bc}) can be written as two singular gauge transformations 
\begin{equation}\Psi(r,\theta=2\pi)=U(\phi)V_{n}(\theta)\Psi(r,\theta=0),\end{equation} 
where $U(\phi)=e^{i\frac{\phi}{2}\sigma_{z}\tau_{z}}$ and $V_{n}(\theta)=e^{i\frac{n\theta}{4}\tau_{y}\sigma_{y}}$.

Let us now consider an additional external magnetic flux $\Phi$ applied at the centre of the defect hole (a magnetic flux tube) corresponding to an azimuthal vector potential
\begin{eqnarray}
\vec{A}=\frac{\Phi}{\Omega_{n}\Phi_{0}r}\hat{\phi}
\end{eqnarray}
with $\Omega_{n}=(1-\frac{n}{6})$ and $\Phi_{0}=\frac{h}{e}$, the magnetic flux quantum.
We write the Hamiltonian in Eq.~(\ref{Low-Hamil}) in polar coordinate and apply the gauge transformation through $U(\phi)$ and $V_n(\theta)$. The transformed Hamiltonian is denoted by $\tilde H(r,\theta)=U^{\dagger}V_n^{\dagger}H V_n U$,
\begin{eqnarray}
\tilde H(r,\theta)&=&\hbar v_{F}(k_{r}-\frac{i}{2r})\tau_{z} \sigma_{x}
\nonumber\\&&
+\hbar v_{F}(k_{\theta}+\frac{\Phi}{\Omega_{n}\Phi_{0}r}+
\frac{n}{4\Omega_{n}r}\tau_{z})\sigma_{y}
+\Delta\tau_{z} \sigma_{z}\quad
\label{Hamiltonian}
\end{eqnarray}
where $k_{r}=-i{\partial \over \partial r}$ and $k_{\theta}=-\frac{i}{r\Omega_{n}}{\partial \over \partial \theta}$.  Written in matrix form it is
\begin{widetext}
\begin{eqnarray}
\tilde H(r,\theta)=
\begin{pmatrix}
\tau \Delta & -i(\partial_{r}+\frac{1}{2r})-\frac{1}{r\Omega_{n}}(\partial_{\theta}+i\frac{\Phi}{\Phi_{0}}+i\frac{n}{4}\tau)\\
-i(\partial_{r}+\frac{1}{2r})+\frac{1}{r\Omega_{n}}(\partial_{\theta}+i\frac{\Phi}{\Phi_{0}}+i\frac{n}{4}\tau) & -\tau \Delta
\end{pmatrix}\label{Hamiltonianmatrix}
\end{eqnarray}
\end{widetext}
where $\tau=+1$ and $-1$ correspond to the $K$ and $K'$ points respectively. 
The eigenstates of Eq.~(\ref{Hamiltonianmatrix}) are two component spinors which, for each valley index, are given  in polar coordinates by
\begin{eqnarray}
\Psi_\tau(r,\theta)=e^{iJ\theta}
\begin{pmatrix}
\Phi_{A}(r)\\
i\Phi_{B}(r)
\end{pmatrix}
\label{eigenf}
\end{eqnarray}
where $J$ is the total angular momentum and $\Phi_{A}(r)$ and $\Phi_{B}(r)$ are linear combinations of Bessel functions due to the cylinder geometry. 
Now we will proceed the calculations by freezing the radial motion. Considering a one-dimensional ring 
and  simplifying the Schr\"odinger equation by discarding the radial variation of electron wavefunction is  a correct approximation for a narrow ring for which the dynamics is frozen in the lowest radial mode. Excited radial modes have an energy much higher and can be neglected~\cite{Fertig} when one studies low energy properties. 
The radial part of the momentum is thus set to zero ($P_{\rho}=-i\hbar({\partial \over \partial \rho}+\frac{1}{2\rho})\to 0$ in matrix elements~\cite{Morpurgo,BCM}). 
In the case of a ring of fixed radius $r=\rho$, this leads to the substitution ${\partial \over \partial r} \rightarrow -\frac{1}{2\rho}$ in Eq.~(\ref{Hamiltonian}) and the eigenstate (\ref{eigenf}) becomes
\begin{eqnarray}
\Psi_\tau(r=\rho,\theta)=e^{iJ\theta}
\begin{pmatrix}
\Phi_{A}(\rho)\\
i\Phi_{B}(\rho)
\end{pmatrix}
\label{eigenf1}
\end{eqnarray}
where $\Phi_A(\rho)$ and $\Phi_B(\rho)$ are now fixed amplitudes.
Due to cylindrical symmetry, the spinors (\ref{eigenf1}) are eigenfunctions of the total angular momentum $J_{z}=L_{z}+ S_{z}$ with $S_{z}=\frac{1}{2}\hbar\sigma_{z}$ and $\hbar(m+\frac{1}{2})$ is the eigenvalue of total angular momentum ($m\in\mathbb{Z}$). People have often used different notations for the wave function \cite{Farias,Farias_Erratum,Romera} like
\begin{eqnarray}
\Psi_\tau(\rho,\theta)=
\begin{pmatrix}
\Phi_{A}(\rho)e^{im\theta}\\
i\Phi_{B}(\rho)e^{i(m+1)\theta}
\end{pmatrix}
\label{eigenf2}
\end{eqnarray}
The wavefunction in Eq~(\ref{eigenf2}) can be obtained from Eq.~(\ref{eigenf1}) by the unitary transformation $U=e^{-i\sigma_{z}\theta/2}$. So, the physical results should  remain unchanged while using any of the the two wavefunctions. In the following our calculations are always consistent in the limit $n=0$ (no disclination) with the results obtained in the literature\cite{Farias,Farias_Erratum,Romera,BMB}.

\subsection{Eigenvalues and Eigenfunctions}
By writing now explicitly $\tilde H\Psi_\tau(\rho,\theta)=E\Psi_\tau(\rho,\theta)$, we obtain
\begin{eqnarray}
\left[\frac{1}{\Omega_{n}}\left(J+\frac{\Phi}{\Phi_{0}}+\frac{n}{4}\tau\right)\right]\Phi_{B}&=&(\mathcal{E}-\tau\delta)\Phi_{A}
\label{eigen1}\\
\left[\frac{1}{\Omega_{n}}\left(J+\frac{\Phi}{\Phi_{0}}+\frac{n}{4}\tau\right)\right]\Phi_{A}&=&(\mathcal{E}+\tau\delta)\Phi_{B}
\label{eigen2}
\end{eqnarray}
with $\mathcal{E}=E/E_{0}$, where $E_{0}=\frac{\hbar v_{F}}{\rho}$ and $\delta=\frac{\Delta}{E_{0}}$. The energy spectrum is given by solving Eqs.~(\ref{eigen1}) and (\ref{eigen2})
\begin{eqnarray}
\mathcal{E}^{\tau}_{n,J}=s\sqrt{\delta^2+\frac{1}{\Omega^{2}_{n}}\left(J+\beta+\frac{n}{4}\tau\right)^2}
\label{energy1}
\end{eqnarray}
which may be also written as
\begin{widetext}
\begin{eqnarray}
\mathcal{E}^{\tau}_{n,J}=s\sqrt{\delta^2+\frac{1}{\Omega^{2}_{n}}(m+\beta+1+\frac{n}{4}\tau)(m+\beta+\frac{n}{4}\tau)+\frac{1}{4\Omega^{2}_{n}}}
\label{energy}
\end{eqnarray}
\end{widetext}
where we have defined the reduced magnetic flux $\beta=\frac{\Phi}{\Phi_{0}}$ and the labels $s=+1$ and $s=-1$ which refer to electron and hole bands respectively. The energy spectra for $K$ and $K'$ valleys are not same.  The valley degeneracy ($\tau$) is broken by the presence of the disclination. This broken degeneracy has important effects on charge currents and wave packet revival time, which we will discuss later.
The amplitudes, reminiscent of the radial part of two spinor components are
\begin{eqnarray}
\Phi^{\tau}_{A}(\rho)&=&1, \\ \Phi^{\tau}_{B}(\rho)&=&\frac{\nu_{\tau}}{\mathcal{E}+\tau \delta}\equiv \xi_{\tau}
\end{eqnarray}
where we defined, $\nu_{\tau}=\frac{J+\beta+\frac{n}{4}\tau}{\Omega_{n}}$.

\begin{figure}
\rotatebox{0}{\includegraphics[width=2.5in]{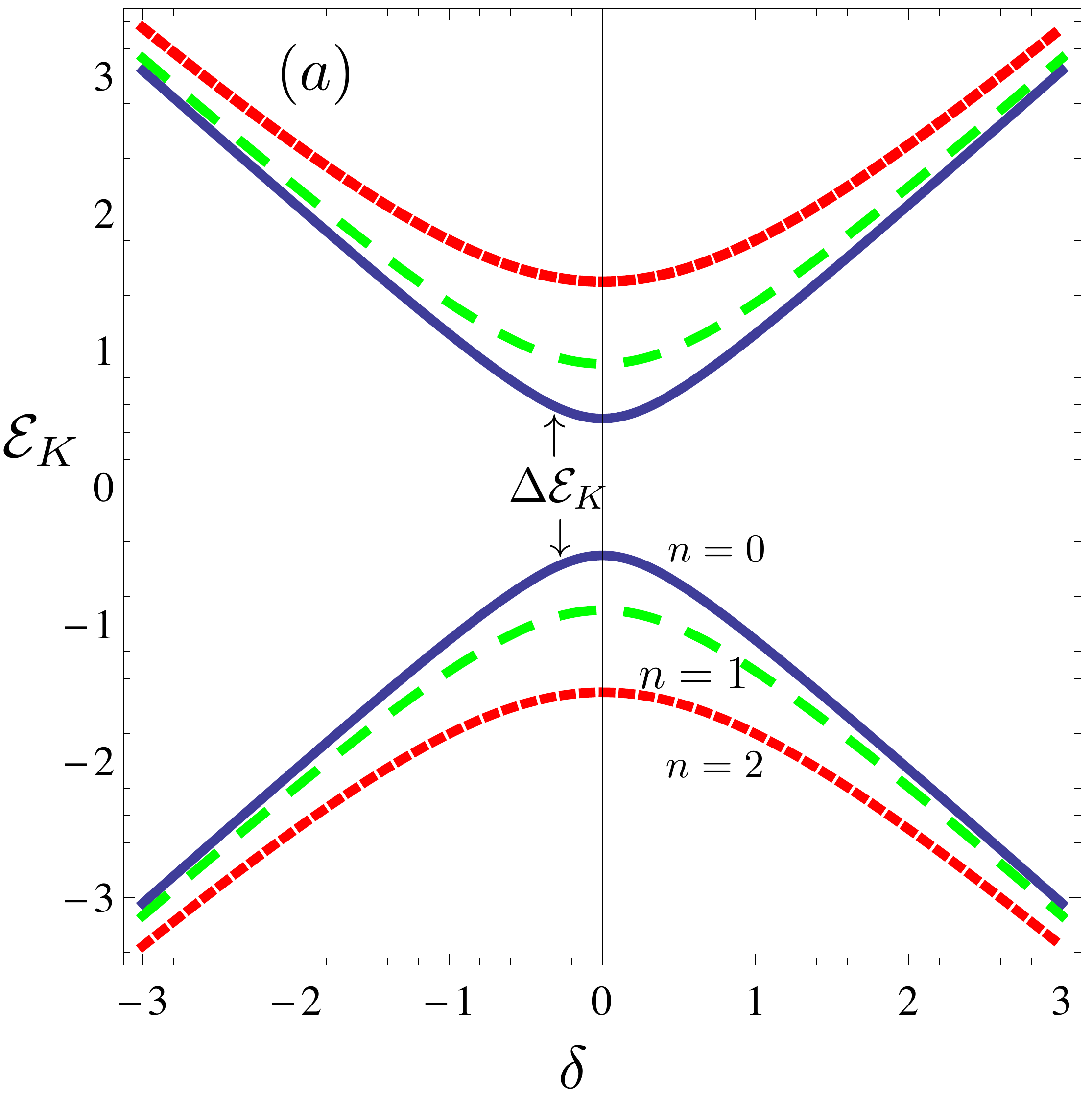}}
\rotatebox{0}{\includegraphics[width=2.5in]{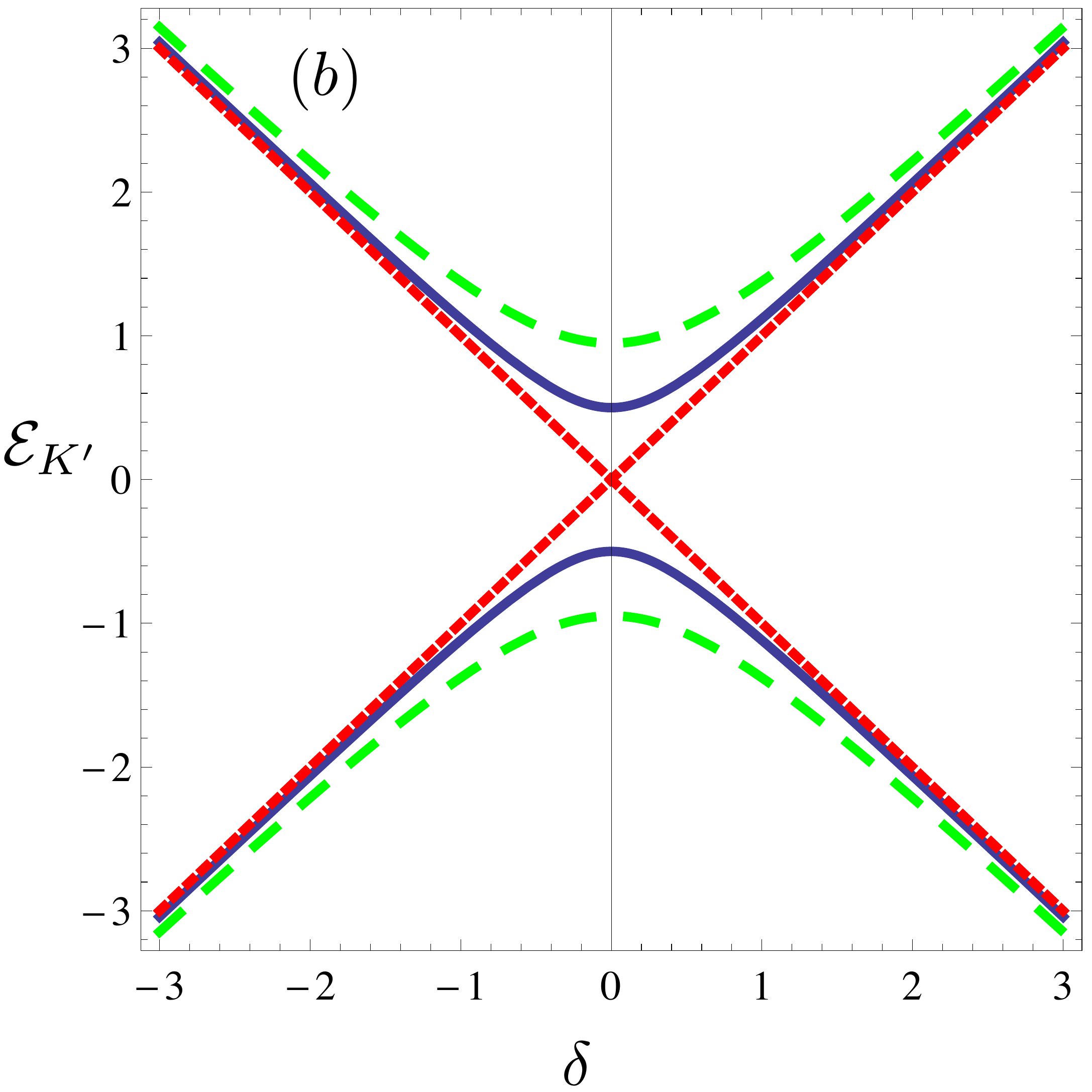}}
\caption{Energy levels (a) $\mathcal{E}_{K}$ and (b) $\mathcal{E}_{K'}$ of a single layer graphene quantum ring as a function of a mass term $\delta$ with $\beta=0$ and $m=0$. Blue, green and red curves are for graphene with $n=0,1,2$ respectively.}
\label{Fig1}
\end{figure}

Here we fix the parameters $\Delta=20$~meV and $\rho=50$~nm. Fig~\ref{Fig1} shows the energy level (for reasons of clarity, in the legends of figures we will denote  physical quantities with $K$ or $K'$ for the two Dirac points (like in  $\mathcal{E}_{K}$) instead of the valley index $\tau$) of a graphene ring for one valley, $\tau=+1$, as a function of the rescaled substrate interaction $\delta$ in absence of magnetic field. From Eq.~(\ref{energy}), the energy difference between the valence and conduction bands is estimated to  $\Delta\mathcal{E}^{\tau}_{n,J}=\frac{2}{\Omega_{n}}(J+\beta+\frac{n}{4}\tau)$ for the value of $\delta=0$. So, the energy difference between two levels depends on the value of the disclination  index $n$ for a given value of quantum number $J$. It  also takes different values  at $K$ and $K'$  points. Interesting features can thus appear. For example, at the $K$ point, the graphene sheet with square defects is gapped (insulator) but at the other Dirac point $K'$ it is gapless (metallic).

In presence of a  finite value of magnetic flux, the system becomes gapless for a fixed value of $\beta=-(J+\frac{n}{4}\tau)$ i.e. $\Phi=-(m+\frac{1}{2}+\frac{n}{4}\tau)\Phi_{0}$ (for $\delta=0$), otherwise the system remains gapped. For example, for $m=0$ the graphene sheet with a pentagonal defect becomes gapless for $\Phi=-\frac{3}{4}\Phi_{0}$ and $\Phi=-\frac{1}{4}\Phi_{0}$ at $K$ and $K'$ points respectively. Similarly, a graphene layer with a defect of index $n=2$ becomes gapless for $\Phi=-\Phi_{0}$ and $\Phi=0$ at $K$ and $K'$ points respectively. Graphene with a negative curvature ($n$ is negative) exhibits the opposite effect. As for example, graphene with $n=-1$ becomes gapless for $\Phi=-\frac{1}{4}\Phi_{0}$ and $\Phi=-\frac{3}{4}\Phi_{0}$ at $K$ and $K'$ points respectively. 

\begin{figure}
\rotatebox{0}{\includegraphics[width=2.5in]{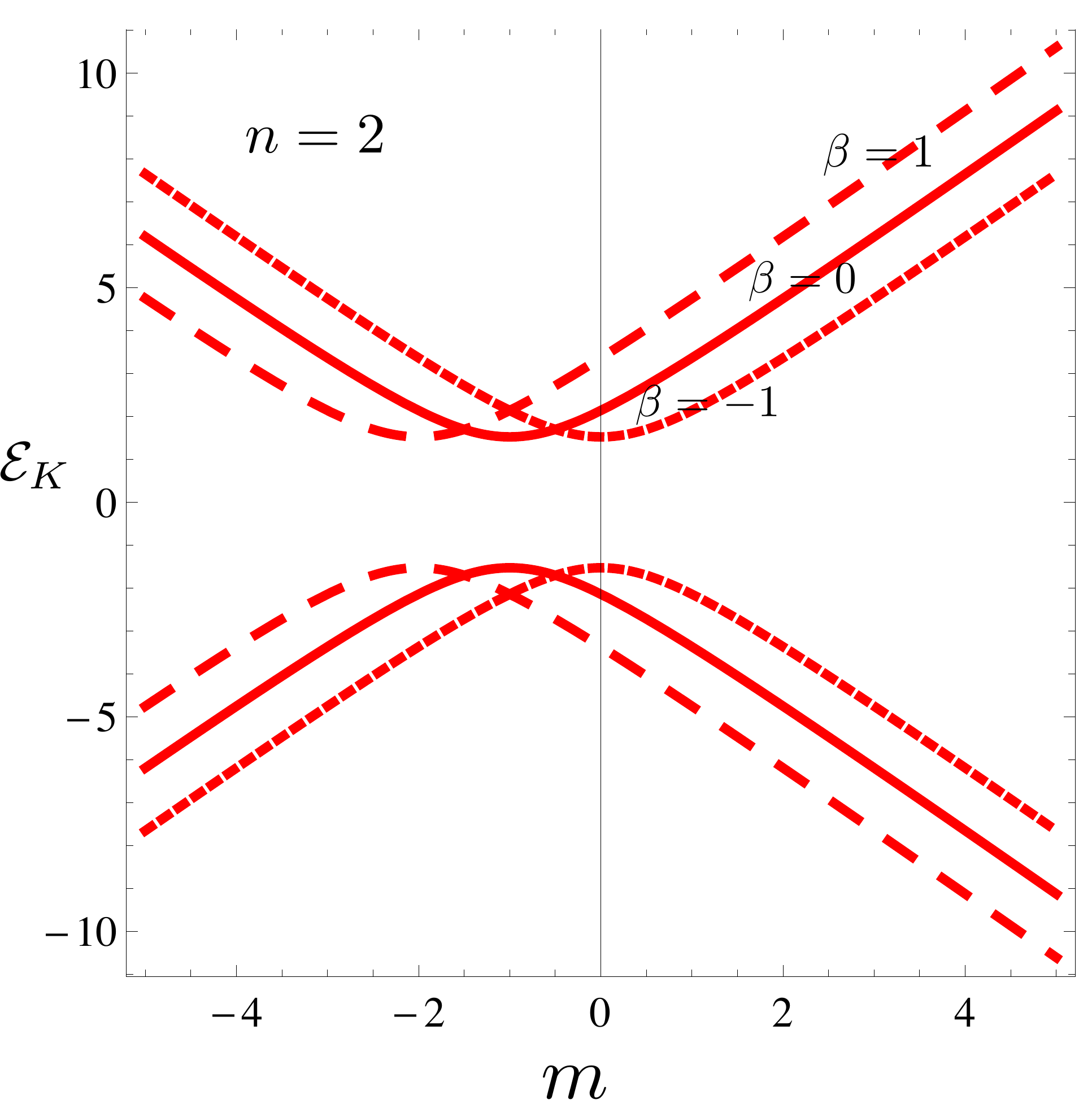}}
\caption{Energy levels $\mathcal{E}_{K}$ for a graphene layer with a square defect as a function of $m$ for different values of the reduced magnetic flux $\beta$. The solid curve is for $\beta=0$, large dashed  curve for $\beta=1.0$ and small dashed curve for $\beta=-1.0$.}
\label{eng_ang}
\end{figure}

In Fig.~\ref{eng_ang} we present the results for the energy levels as a function of the angular momentum $m$ for $\beta=0$, $1.0$ and $-1.0$ respectively. From Eq.~(\ref{energy}), the energy spectrum exhibits a minimum for $m=-(\frac{1}{2}+\beta+\frac{n}{4}\tau)$ for a given value of $\beta$ and it is independent of $\delta$. So, the gauge field due to curvature shifts the energy minimum and has an opposite effect at $K$ and $K'$ points.

\begin{figure}
\rotatebox{0}{\includegraphics[width=2.5in]{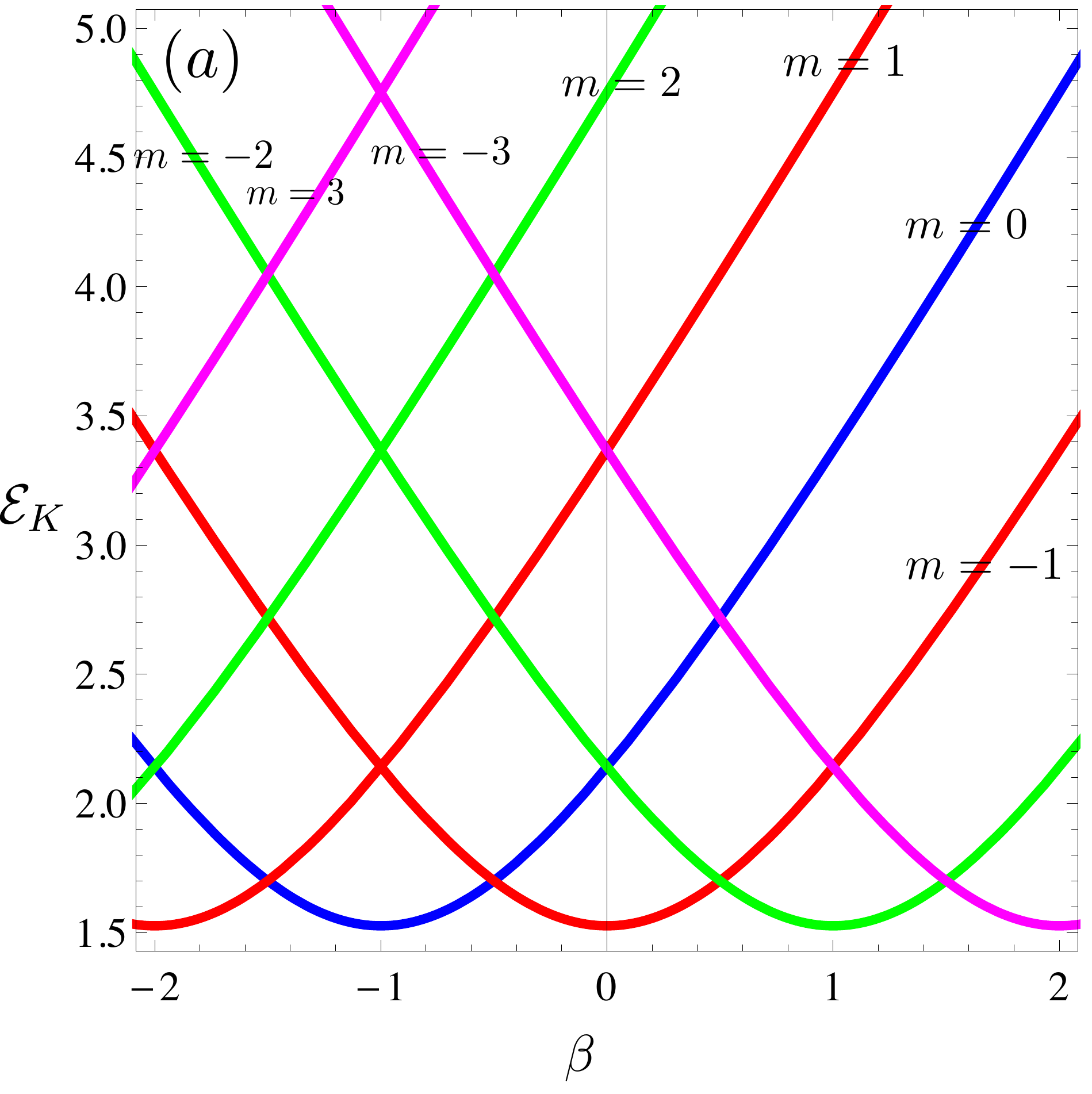}}
\rotatebox{0}{\includegraphics[width=2.5in]{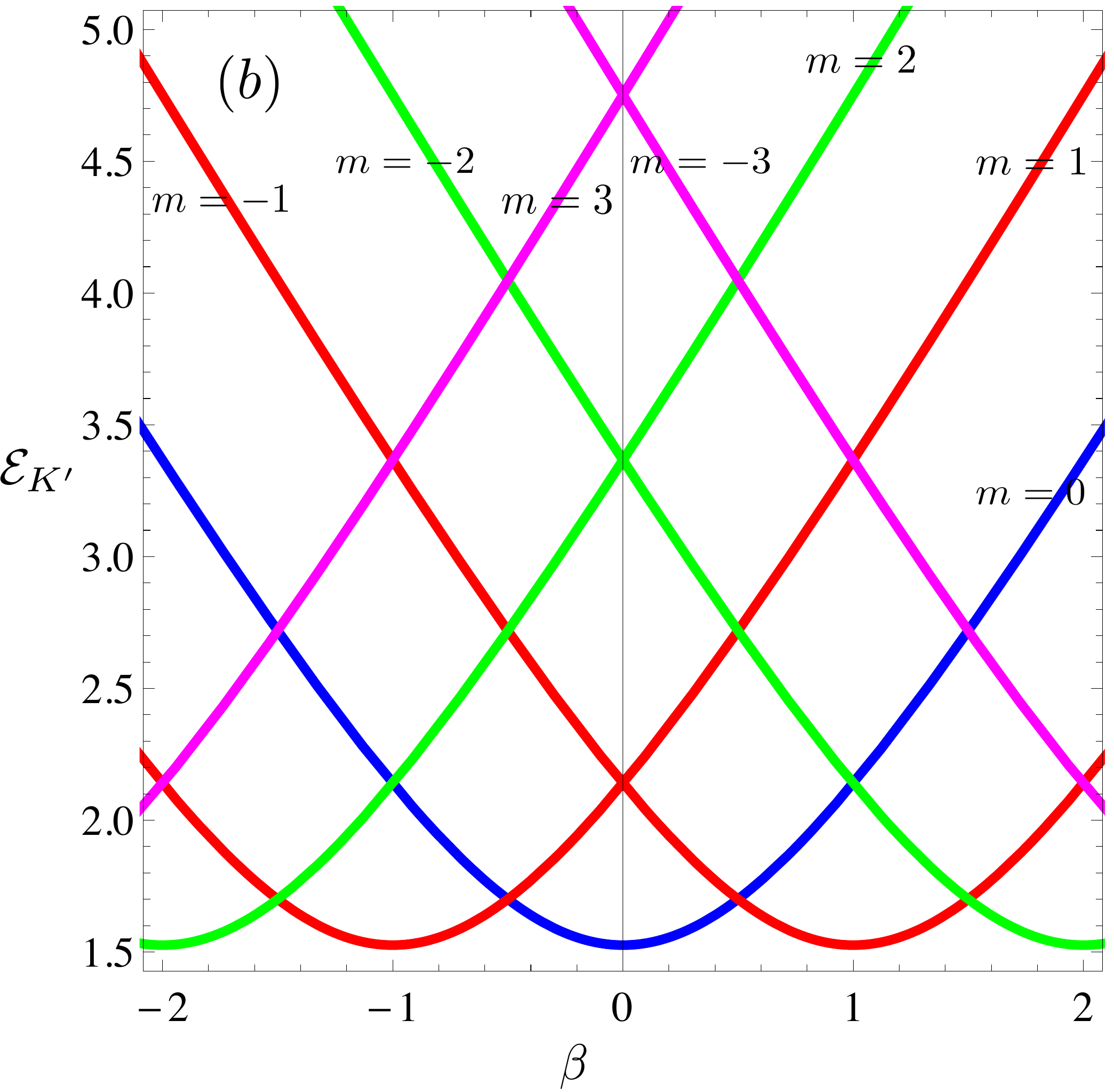}}
\caption{Energy levels (a) $\mathcal{E}_{K}$ and (b) $\mathcal{E}_{K'}$ for a graphene layer with a square defect as a function of $\beta=\frac{\Phi}{\Phi_{0}}$. The blue curve is for $m=0$, red for $m=-1$ and $m=1$, green for $m=-2$ and $m=2$, magenta for $m=-3$ and $m=3$, respectively.}
\label{eng_mag_sq}
\end{figure}

\begin{figure}
\rotatebox{0}{\includegraphics[width=2.5in]{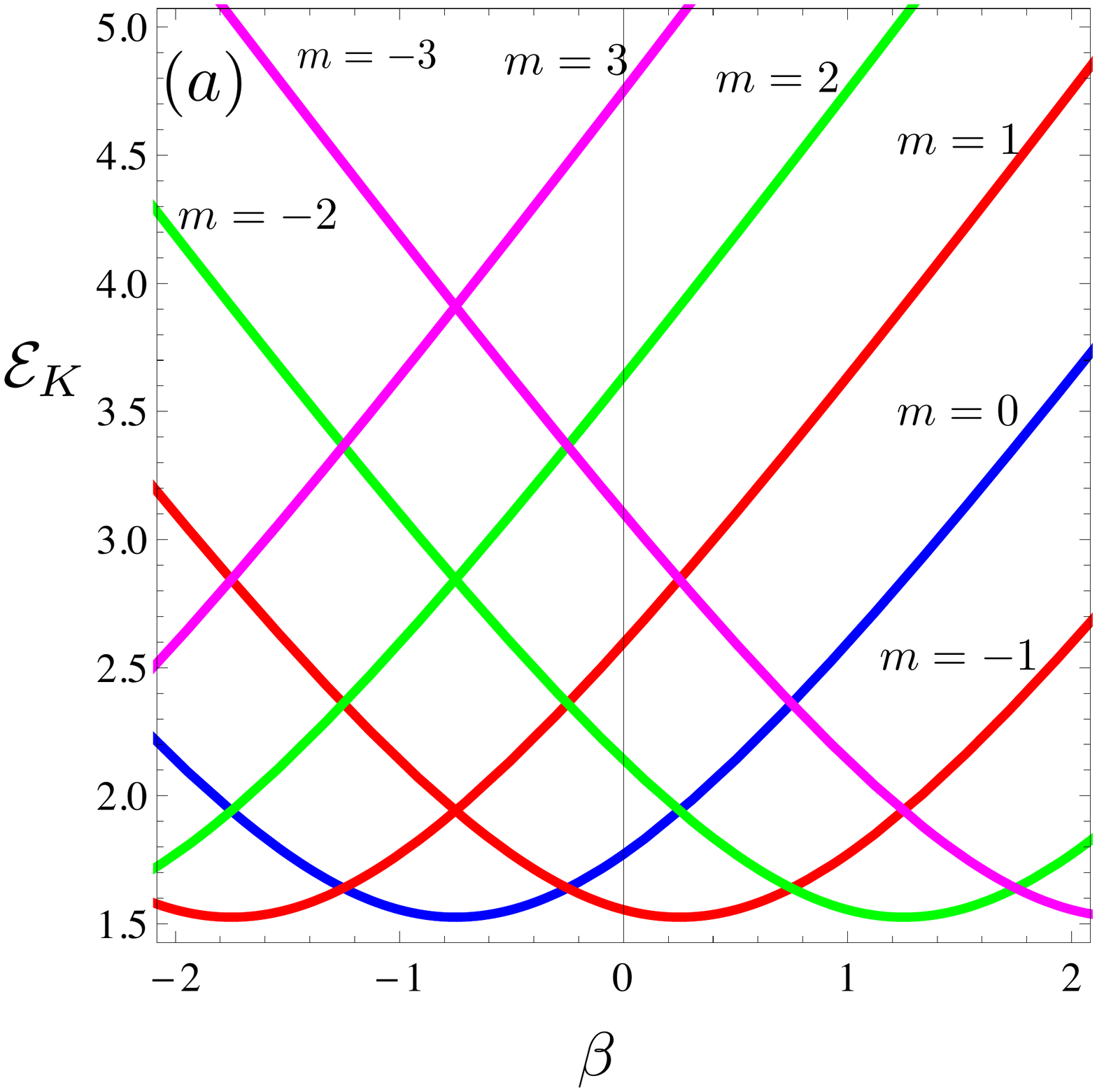}}\vspace{-10mm}
\rotatebox{0}{\includegraphics[width=2.5in]{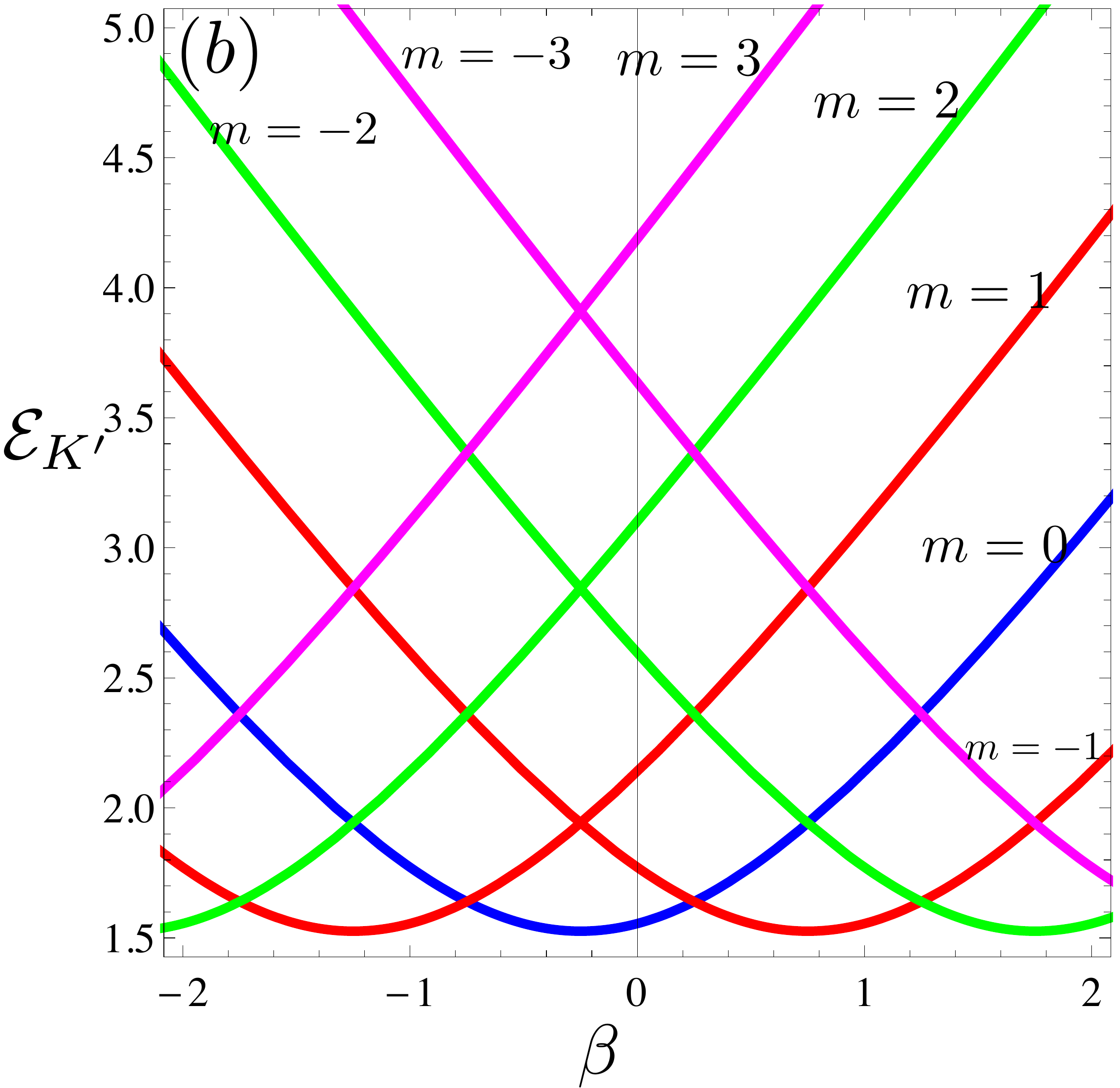}}
\caption{Energy levels (a) $\mathcal{E}_{K}$ and (b) $\mathcal{E}_{K'}$ for a pentagonal defect in a  graphene layer as a function of $\beta=\frac{\Phi}{\Phi_{0}}$. The color significance is the same as in Fig.(\ref{eng_mag_sq})}
\label{pent_mag}
\end{figure}

In Figs~(\ref{eng_mag_sq}) and (\ref{pent_mag}) we plot the energy as a function of the reduced magnetic flux ($\beta$) for different values of $m$.

\subsection{Pseudo-spin polarization}
The pseudo-spin polarization is a contribution to the orbital angular momentum~\cite{Schliemann,Mecklenburg}. It can be calculated from the eigenvectors. The components of the pseudo-spin polarization are given by 
\begin{eqnarray}
\<\sigma_{x}\>&=&0\nonumber\\
\<\sigma_{y}\>&=&\frac{2\xi_{\tau}}{1+{\xi_{\tau}}^{2}}\nonumber\\
\<\sigma_{z}\>&=&\frac{1-{\xi_{\tau}}^{2}}{1+{\xi_{\tau}}^{2}}
\end{eqnarray}
and are 
such that ${\<\sigma_{x}\>}^{2}+\<\sigma_{y}\>^{2}+\<\sigma_{z}\>^{2}=1$.
\begin{figure}
\rotatebox{0}{\includegraphics[width=3in]{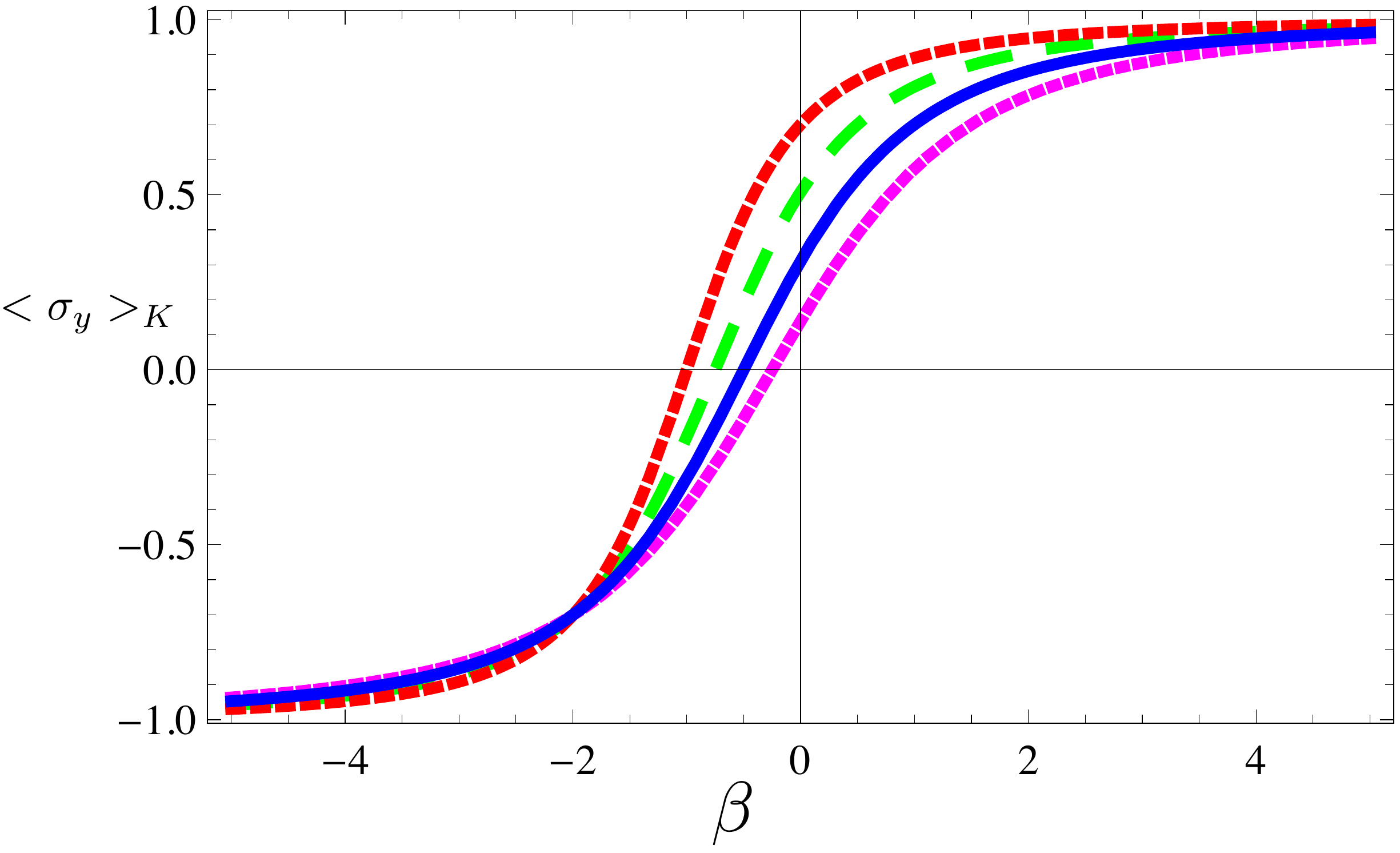}}
\caption{Pseudo-spin polarization $\<\sigma_{y}\>$ as a function of $\beta$ for $m=0$ at point K. The red, green, blue and magenta curve are for graphene with $n=2,1,0,-1$ respectively.}
\label{Spin_Pol}
\end{figure}
In Fig.~\ref{Spin_Pol} we show the curvature effect on pseudo-spin polarization ($\sigma_{y}$) of a disclinated graphene. It is seen that at $\beta=0$, positive curvature has a larger spin-polarization. The blue curve in Fig.~\ref{Spin_Pol} is for a monolayer graphene ring. Due to the ring geometry it also has a permanent pseudo-spin polarization at $\beta=0$.

\subsection{Charge Currents}
The angular velocity is defined according to $v_{\theta}=\frac{i\rho}{\hbar}[\tilde H(\rho,\theta),\theta]=\frac{1}{\Omega_{n}}\sigma_{y}$. So, the current is given by
\begin{eqnarray}
J_{\theta}=\frac{1}{\Omega_{n}}({\Phi^{\tau}_{A}}^{\star}\Phi^{\tau}_{B}+\Phi^{\tau}_{A}{\Phi^{\tau}_{B}}^{\star})=\frac{2}{\Omega_{n}}\xi_{\tau}
\end{eqnarray}

\begin{figure}
\rotatebox{0}{\includegraphics[width=3in]{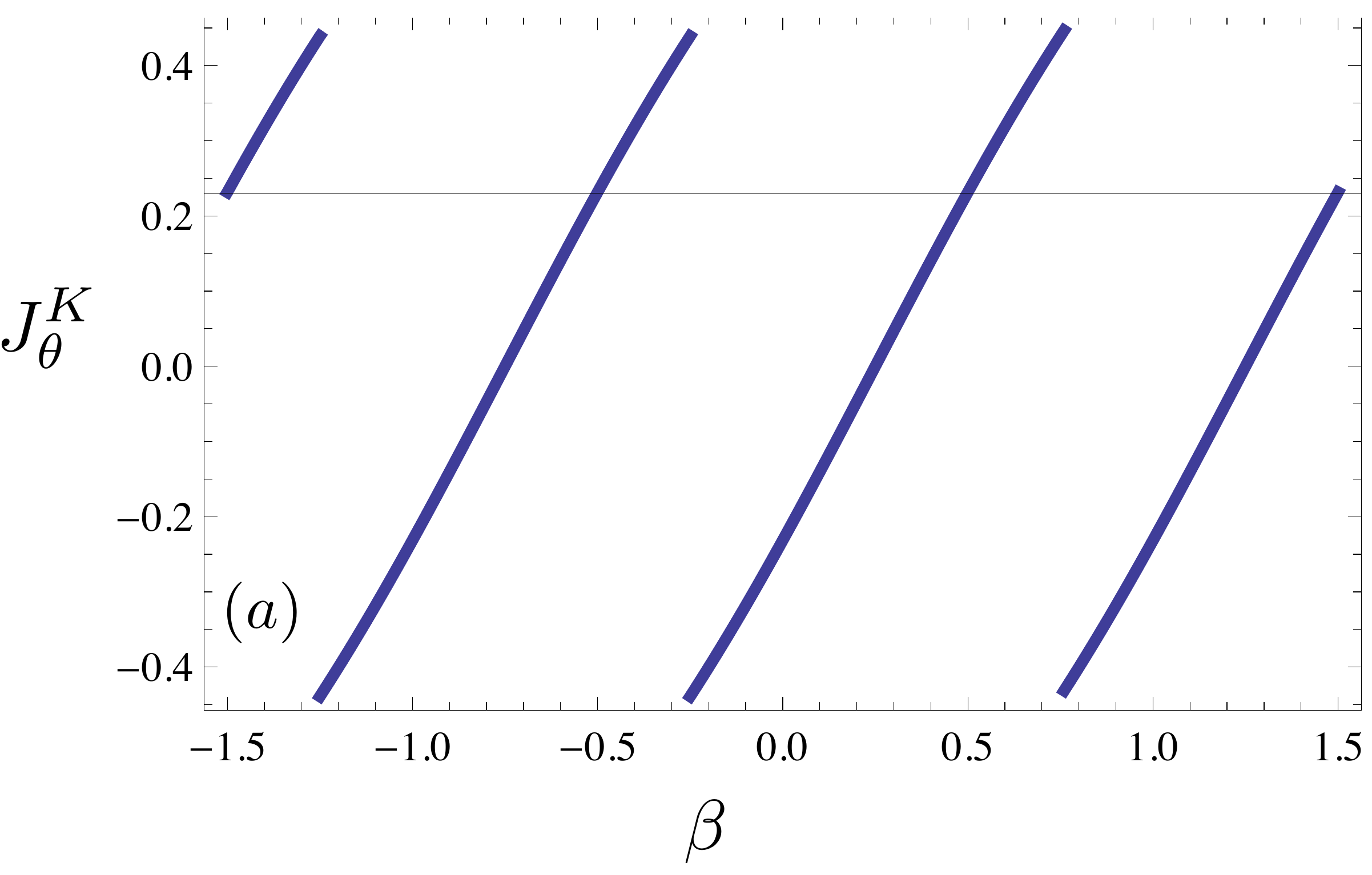}}
\rotatebox{0}{\includegraphics[width=3in]{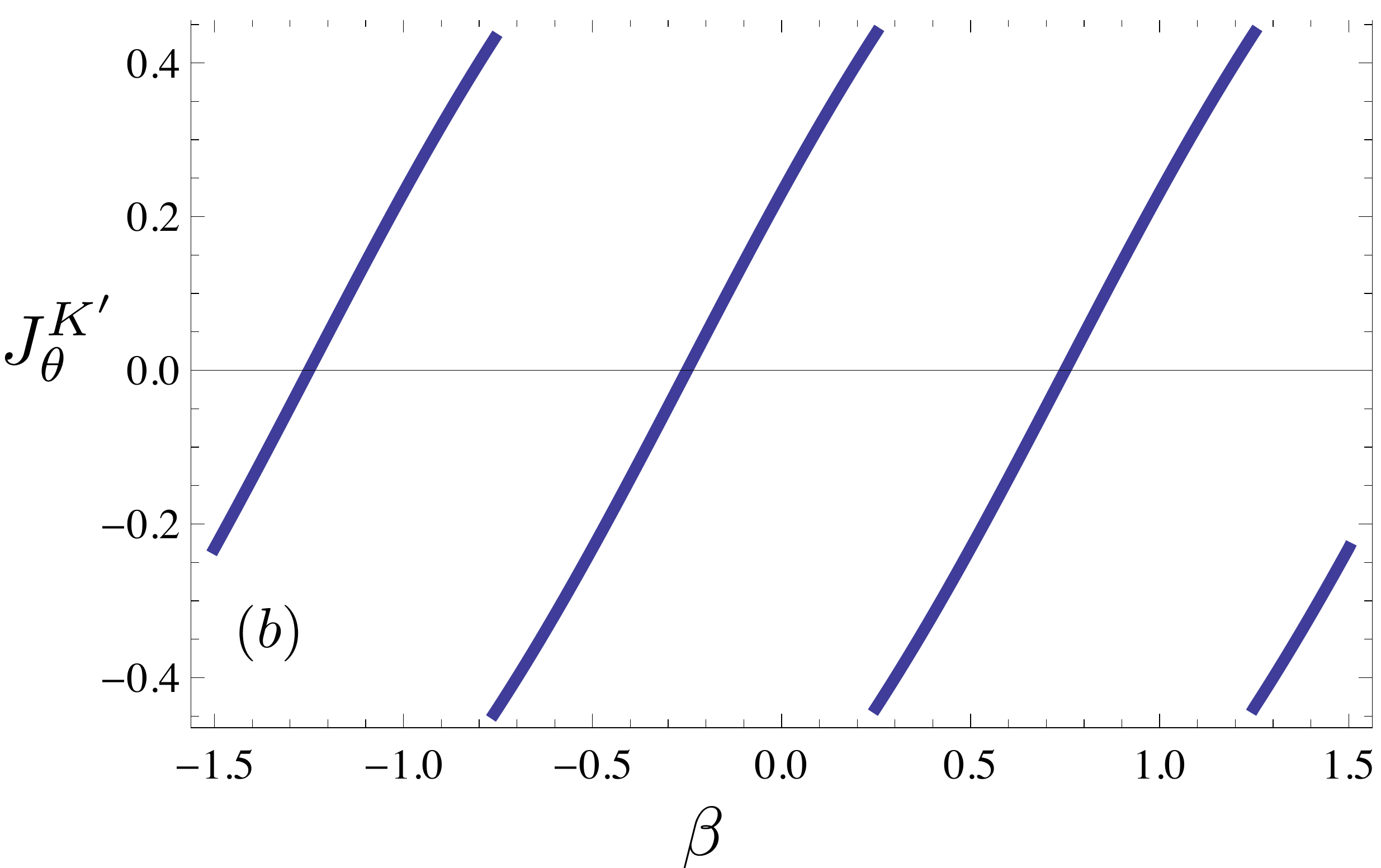}}
\rotatebox{0}{\includegraphics[width=3in]{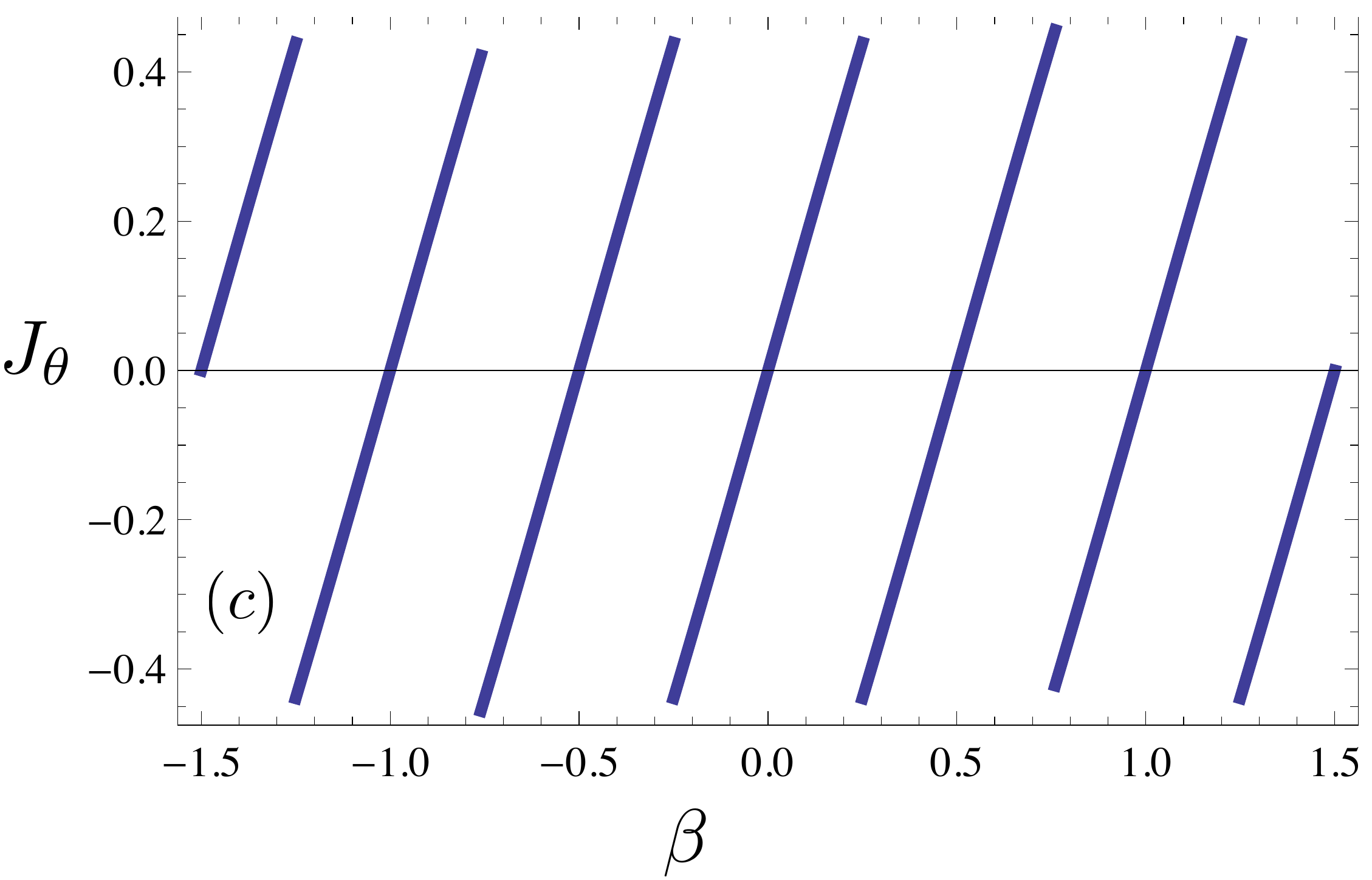}}
\caption{Charge current $J_{\theta}$ for a pentagon graphene ring. Fig(a) shows the charge current for the valley $\tau=+1$ and fig(b) for valley $\tau=-1$. Fig(c) is for the total charge currents.}
\label{pent_current}
\end{figure}

\begin{figure}
\rotatebox{0}{\includegraphics[width=3in]{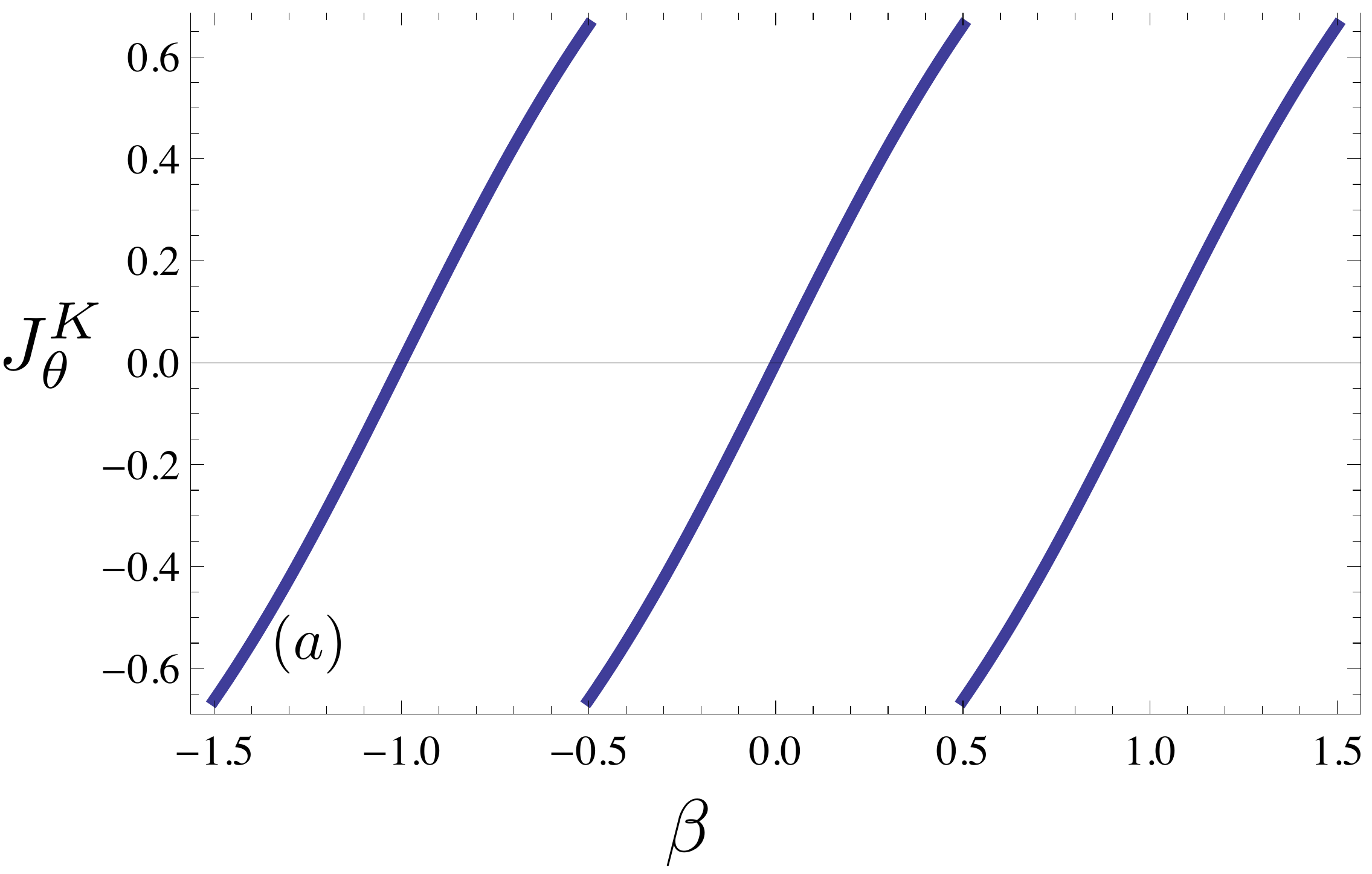}}
\rotatebox{0}{\includegraphics[width=3in]{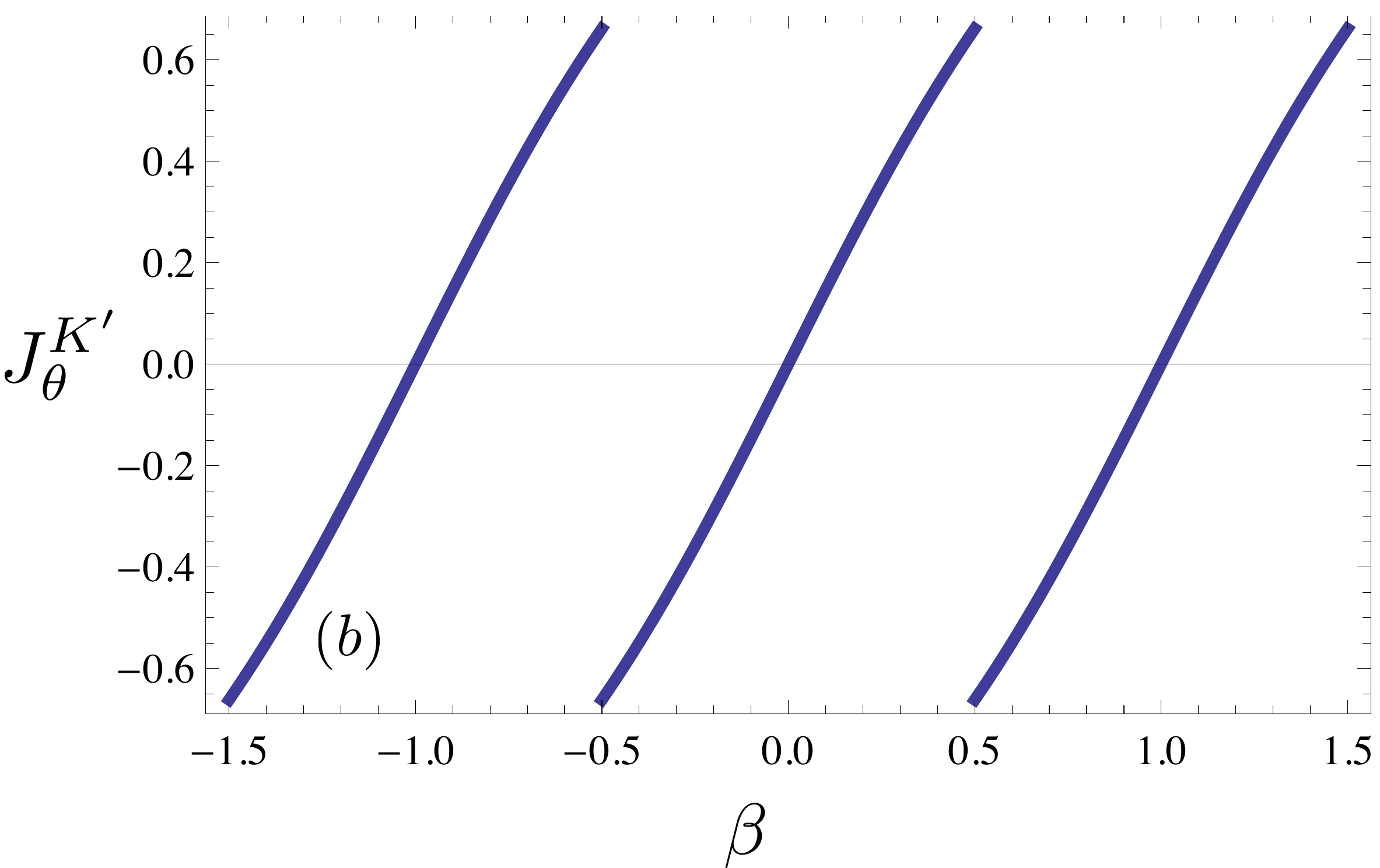}}
\rotatebox{0}{\includegraphics[width=3in]{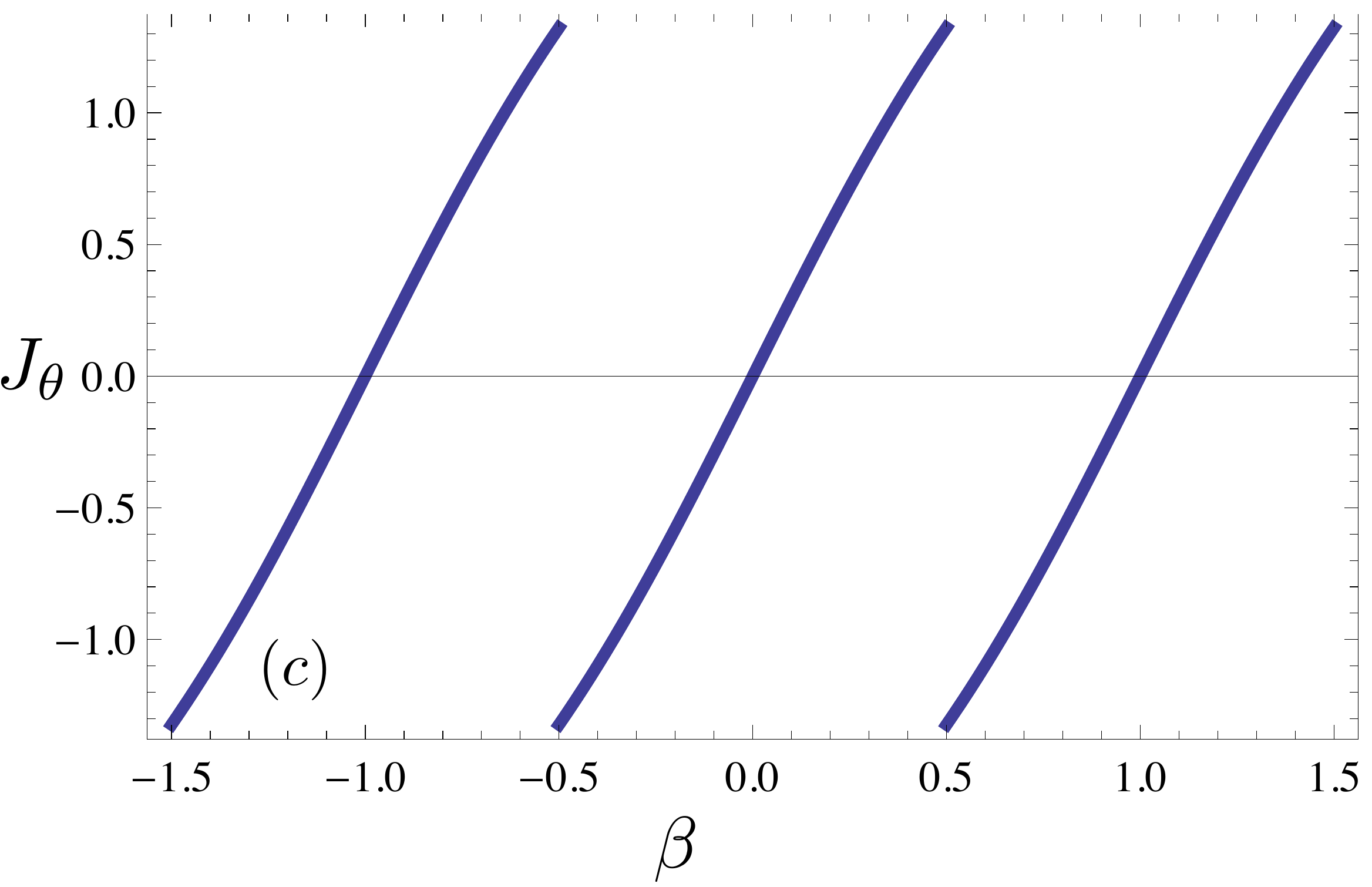}}
\caption{Charge current $J_{\theta}$ for a graphene ring with square defect for $\tau=+1$ and $\tau=-1$ in fig(a) and fig(b) respectively. The total charge current shows in fig(c).}
\label{sq_current}
\end{figure}

The  expressions of the angular current in valleys $K$ and $K'$ are found to be
\begin{eqnarray}
J^{K}_{\theta}=\frac{2}{\Omega^{2}_{n}}\frac{(J+\beta+\frac{n}{4})}{\mathcal{E}_{K}+\delta},
\end{eqnarray}
and
\begin{eqnarray}
J^{K'}_{\theta}=\frac{2}{\Omega^{2}_{n}}\frac{(J+\beta-\frac{n}{4})}{\mathcal{E}_{K'}-\delta}.
\end{eqnarray}
We observe that there are three contributions, first a term due to the orbital angular momentum $(J)$, second a term due to the presence of the magnetic flux $(\beta$) and finally the defect contribution ($n$). It is worth mentioning that the defect contributions are opposite in the two valleys.
The total current is now
\begin{eqnarray}
J_{\theta}&=&J^{K}_{\theta}+J^{K'}_{\theta}\nonumber\\
&=&\frac{2}{\Omega^{2}_{n}}\frac{(J+\beta)(\mathcal{E}_{K}+\mathcal{E}_{K'})-\frac{n}{4}[(\mathcal{E}_{K}-\mathcal{E}_{K'})-2\delta]}{(\mathcal{E}_{K}+\delta)(\mathcal{E}_{K'}-\delta)}
\label{total_current}
\end{eqnarray}
In Figs.~\ref{pent_current} and \ref{sq_current} we have shown the charge current of a graphene  layer with defects of indices $n=1$ and $n=2$. Note that the current contribution from $K$ and $K'$ are not same for pentagon graphene. {This is also true for odd $n$. For even $n$, the contribution from the $K-$valley and $K'$ are the same and they oscillate in phase, this is also true for $n=0$ \cite{Farias_Erratum}. The magnetic flux periodicity of the charge current depends on the index $n$}.

\section{Revival and classical periodicity}\label{Wave Packet Revival}
The revival phenomenon has attracted broad interest over the past decade. Revivals, and associated fractional revivals, are investigated theoretically and also observed experimentally in Rydberg wave packets in atoms, molecules, molecular vibrate states and Bose-Einstein 
condensates\cite{Revival1,Revival2,Revival3,Revival4}.  
The revival method is used to control the wave packet\cite{wave_control1,wave_control2,wave_control3}, or for isotope separation \cite{isotope1,isotope2,isotope3} to give examples of applications. Revivals happen when the wave packet comes back to its initial shape during  its temporal evolution. The time at which the revival occurs is called revival time ($T_{R}$). The periodicity of the wave packet revivals depends on the energy eigenvalue spectrum and is independent of the initial shape of the wave function. For an initial wave packet, written as a superposition of eigenstates sharply peaked around some level $n'_{0}$, the Taylor expansion of the energy spectrum $E_{n'}$ around the the energy $E_{n_{0}}$, is given by
\begin{eqnarray}
E_{n'}\simeq E_{n'_{0}}+E'_{n'_{0}}(n'-n'_{0})+\frac{E''_{n'_{0}}}{2}(n'-n'_{0})^{2}+\dots
\label{Taylor}
\end{eqnarray}
where every coefficient of the Taylor expansion in Eq.~(\ref{Taylor}) gives an important characteristic time scale for the propagating wave packet\cite{Averbukh, Viktor}. The temporal evolution of the localized bound state $\Psi$ for a time independent Hamiltonian can be written in terms of the eigenfunctions $u_{n'}$ and eigenvalues $E_{n'}$ as
\begin{eqnarray}
\Psi=\sum^{\infty}_{n'=0}a_{n'}u_{n'}e^{-iE_{n'}t/\hbar}
\end{eqnarray}
with $a_{n'}=\< {u_{n'}} | \Psi\>$. If the coefficient $a_{n'}$ are considered to be tightly spread around a large $n'_{0}\gg |n'-n'_{0}|$ and $n'_{0}\gg 1$, and taking into account Eq.~(\ref{Taylor}), one has
\begin{eqnarray}
e^{-iE_{n'}t/\hbar}&=&e^{-i/\hbar(E_{n'_{0}}t+E'_{n'_{0}}(n'-n'_{0})+E''_{n'_{0}}/2(n'-n'_{0})^{2}+\dots)}\nonumber\\
&=&e^{-i\omega_{0}t-2\pi i(n'-n'_{0})t/T_{Cl}-2\pi i(n'-n'_{0})^{2}t/T_{R}+\dots}\nonumber\\
\end{eqnarray}
which defines different time scale, that is, $T_{R}=4\pi \hbar/|E''_{n'_{0}}|$ and $T_{Cl}=2\pi \hbar/E'_{n'_{0}}$. The propagating wave initially evolves quasiclassically and oscillates with a period $T_{Cl}$, then it will spread and collapse (delocalize). At later times, given by integral multiples of $T_{R}/2$, the wave packet  will regain its initial shape and oscillate again with a period $T_{Cl}$. Moreover, at times that are rational fractions of $T_{R}$, the wave packet splits into a collection of scaled and reshifted copies called fractional revivals. A longer time scale can be defined beyond $T_{R}$, the so called super-revival time, at which a new cycle of full and fractional revivals emerges again \cite{Robinett}. 

\begin{figure}[th]
\rotatebox{0}{\includegraphics[width=3.2in]{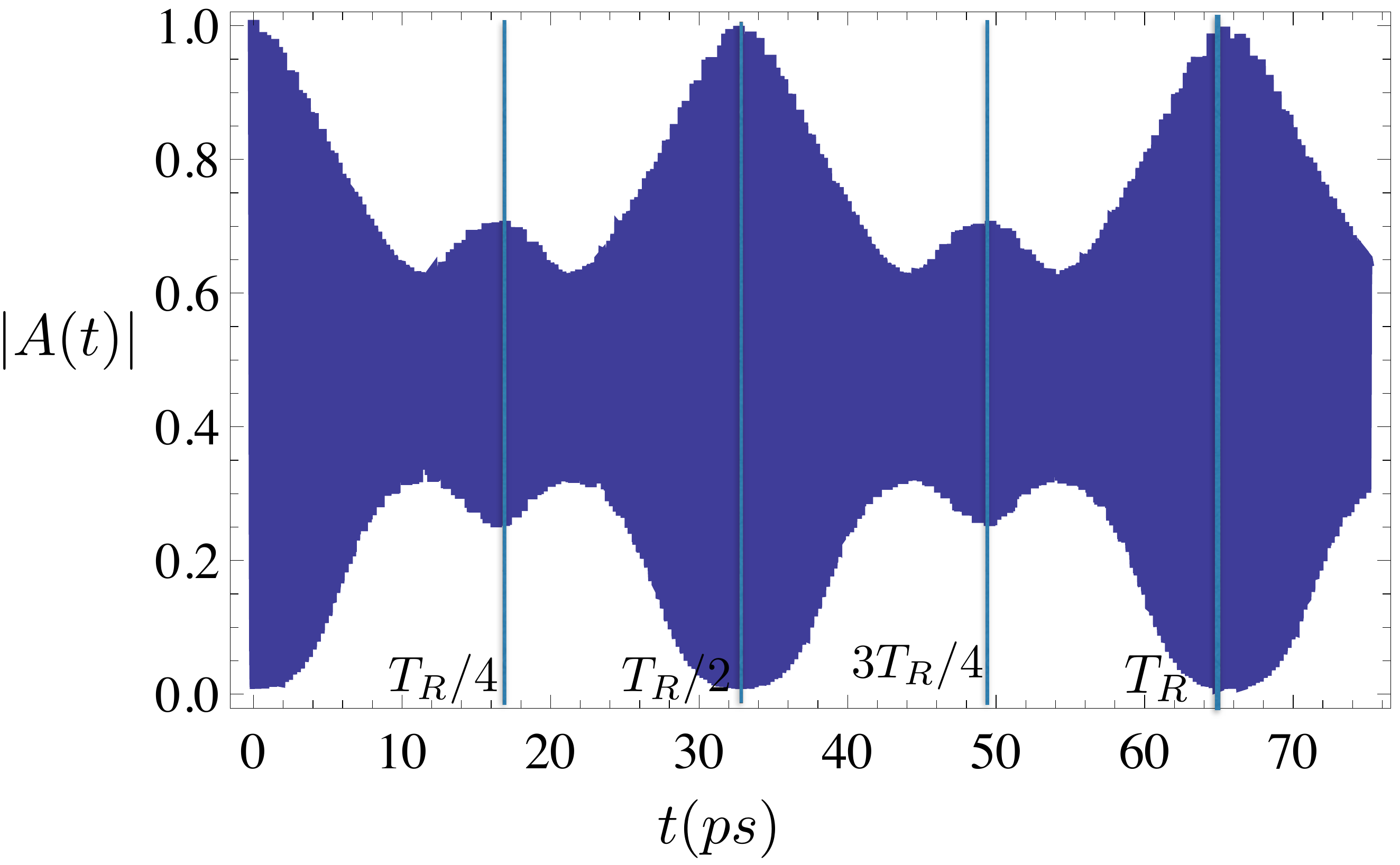}}
\rotatebox{0}{\includegraphics[width=3.2in]{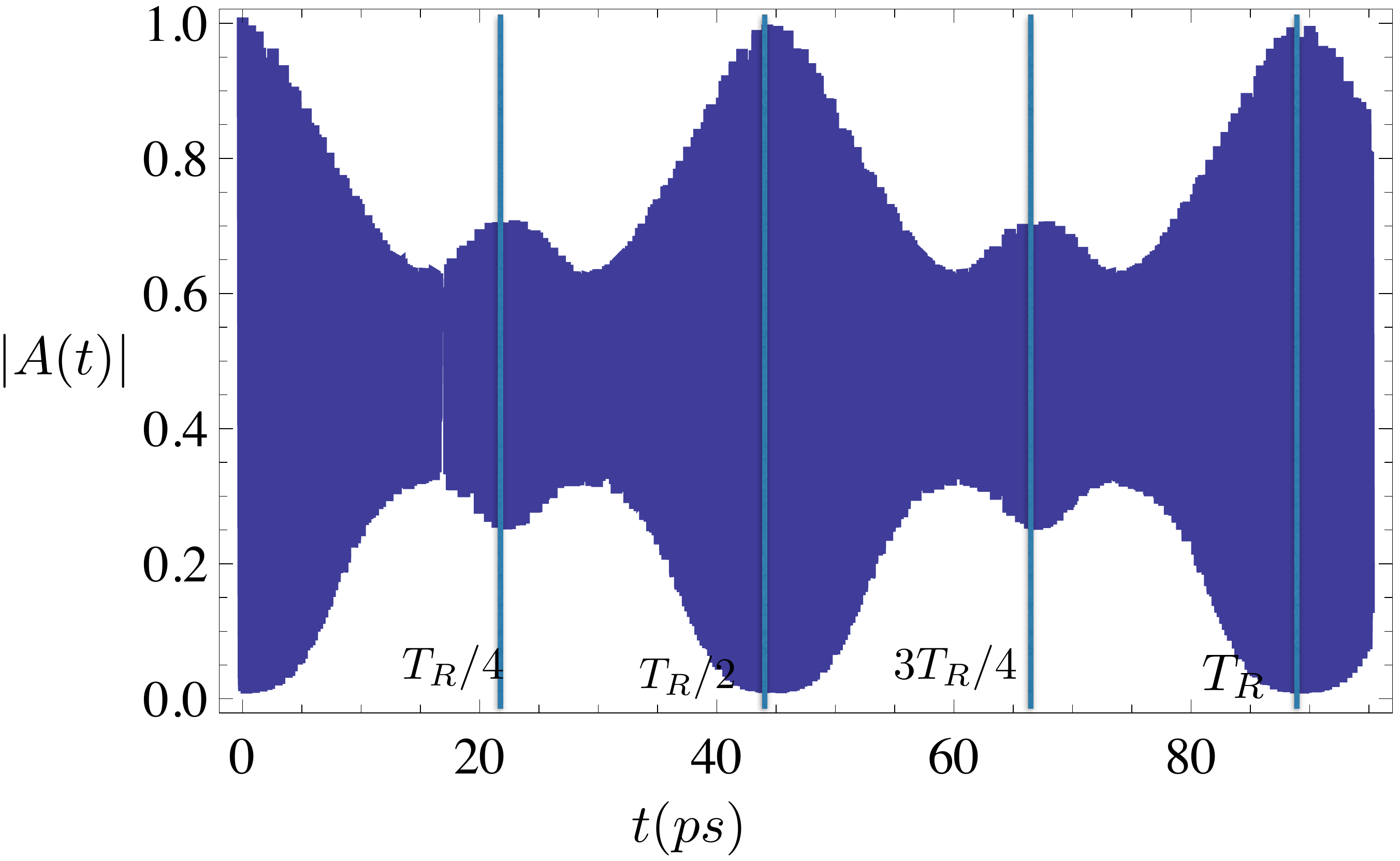}}
\caption{Time dependence of $|A(t)|$ for initial wavepacket with $m=5$ and $\sigma=1.5$ at the $K$ point. Upper pannel: $T_{R}=65.56$~ps with  pentagonal defect. Lower pannel:
$T_{R}=89.35$~ps  with a square defect. The difference lies in the time scales (horizontal axis).}
\label{sq-time}
\end{figure}

We shall construct the initial wavepacket as a linear combination
\begin{eqnarray}
\Psi(\rho,\theta)=\sum_{m}c_{m}\Phi_{m}(\rho,\theta)
\label{initial-wave}
\end{eqnarray}
centered around a given eigenvalue $E_{m_{0}}$, with Gaussian distributed coefficients $c_{m}=\sqrt{\frac{1}{\sqrt{\pi}\sigma}}e^{-\frac{(m-m_{0})^{2}}{2\sigma^{2}}}$, where $\sigma$ is the variance of the distribution. 
We can write the temporal evolution of the initial wavepacket as
\begin{eqnarray}
\Psi_{t}(\rho,\theta)=\sum_{m}\Phi_{m}(\rho,\theta)e^{-iE_{m}t/\hbar}.
\label{time-wave}
\end{eqnarray}
The study of the time dependence of the wavepacket is calculated in terms of the autocorrelation function, that is, the overlap of the initial state $\Psi_{0}$ in Eq.~(\ref{initial-wave}) and its temporal evolution in Eq.~(\ref{time-wave}):
\begin{eqnarray}
A(t)=\< {\Psi_{0}}| {\Psi_{t}}\>=\int^{2\pi}_{0}\Psi_{0}(\rho,\theta)\Psi_{t}(\rho,\theta)d\theta.
\end{eqnarray}
If we consider the expansion of Eq.~(\ref{Taylor}), $A(t)$ can be written as
\begin{eqnarray}
A(t)=\sum_{m}|c_{m}|^{2}e^{-iE_{m}t/\hbar}.
\label{Auto_corr}
\end{eqnarray}
The occurrence of revivals corresponds to a complete overlap with the initial state, for which $|A(t)|^{2}$ returns to its initial value of unity.
In this case the classical period and the revival time can be calculated from Eq.~(\ref{Taylor}) and Eq.~(\ref{energy}) 
\begin{equation}T_{Cl}=\frac{2\pi \rho \Omega_{n}}{v_{F}}\sqrt{1+\frac{\delta^{2}\Omega^{2}_{n}}{(m+\frac{1}{2}+\frac{n}{4}\tau)^{2}}}\end{equation} and 
\begin{equation}T_{R}=\frac{4\pi \rho \Omega^{2}_{n}}{v_{F}}\frac{[\delta^{2}+\frac{1}{\Omega^{2}_{n}}(m+\frac{1}{2}+\frac{n}{4}\tau)^{2}]^{\frac{3}{2}}}{\delta^{2}}\end{equation} 
for $\beta=0$. The characteristic times are depending on the type of disclination and are different at  $K$ and $K'$ points.
In Fig. 
\ref{sq-time} we show plots of the time dependence of $|A(t)|$ for pentagonal and square defects in graphene. In all case we choose $\rho=50$~nm, $\Delta=50$~meV and $m_{0}=5$. For pentagonal graphene the initial wavepacket has a fast quasiclassical oscillatory behavior with a period $T_{Cl}=0.2668$~ps; this quasiclassical amplitude decreases and later there is regeneration of the quasiclassical behaviour at half the revival time $T_{R}/2=32.78$~ps. Similarly, for a square defect $T_{Cl}=0.2133$~ps and $T_{R}=89.35$~ps (see table \ref{table}).

\begin{table}[htb]
\caption{Classical and Revival times at the two isotropic valleys $K$ and $K'$ for a graphene quantum ring with index $n=1$ and $n=2$ with $\rho=50$~nm and $\Delta=20~$meV. The initial wavepacket were built as a superposition of Guassian-distribouted state of width $\sigma=1.5$. }
\vspace{5mm}
\small
\begin{tabular}{|c| c| c| c| c|} 
\hline\hline
$m$ & $T_{Cl}(K)$ & $T_{Cl}(K'),$ & $T_{Cl}(K)$ & $T_{Cl}(K')$ \\ 
 & $n=1$ & $n=1$ & $n=2$ & $n=2$ \\ 
\hline
5 & 0.2668 & 0.2681 & 0.2133 & 0.2146 \\ 
\hline
10 & 0.2624 & 0.2626 & 0.2112 & 0.2114 \\ 
\hline 
15 & 0.2614 & 0.2615 & 0.2108 & 0.2108 \\ 
\hline
20 & 0.2611 & 0.2611 & 0.2106 & 0.2106 \\ 
\hline \hline
$m$ & $T_{R}(K)$ & $T_{R}(K'),$ & $T_{R}(K)$ & $T_{R}(K')$ \\ 
 & $n=1$ & $n=1$ & $n=2$ & $n=2$ \\ 
\hline
5 & 65.56 & 50.68 & 89.35 & 53.20\\
\hline
10 & 407 & 354 & 539 & 406.71\\
\hline
15 & 1268 & 1152 & 1650 & 1360.94\\
\hline
20 & 2888 & 2685 & 3721 & 3216.25\\
\hline
\end{tabular}
\label{table}
\end{table}

\begin{figure}
\rotatebox{0}{\includegraphics[width=3.2in]{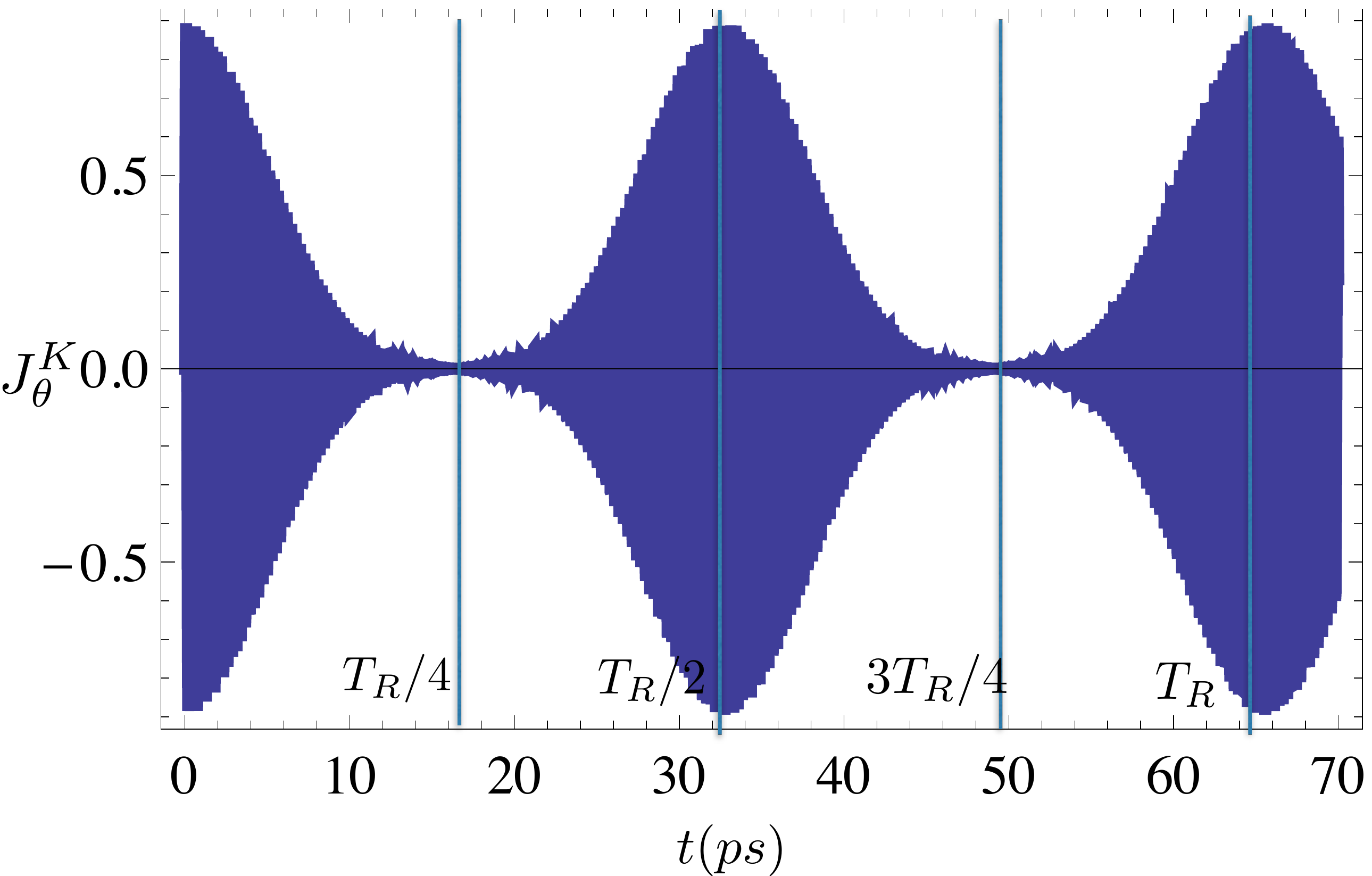}}
\rotatebox{0}{\includegraphics[width=3.2in]{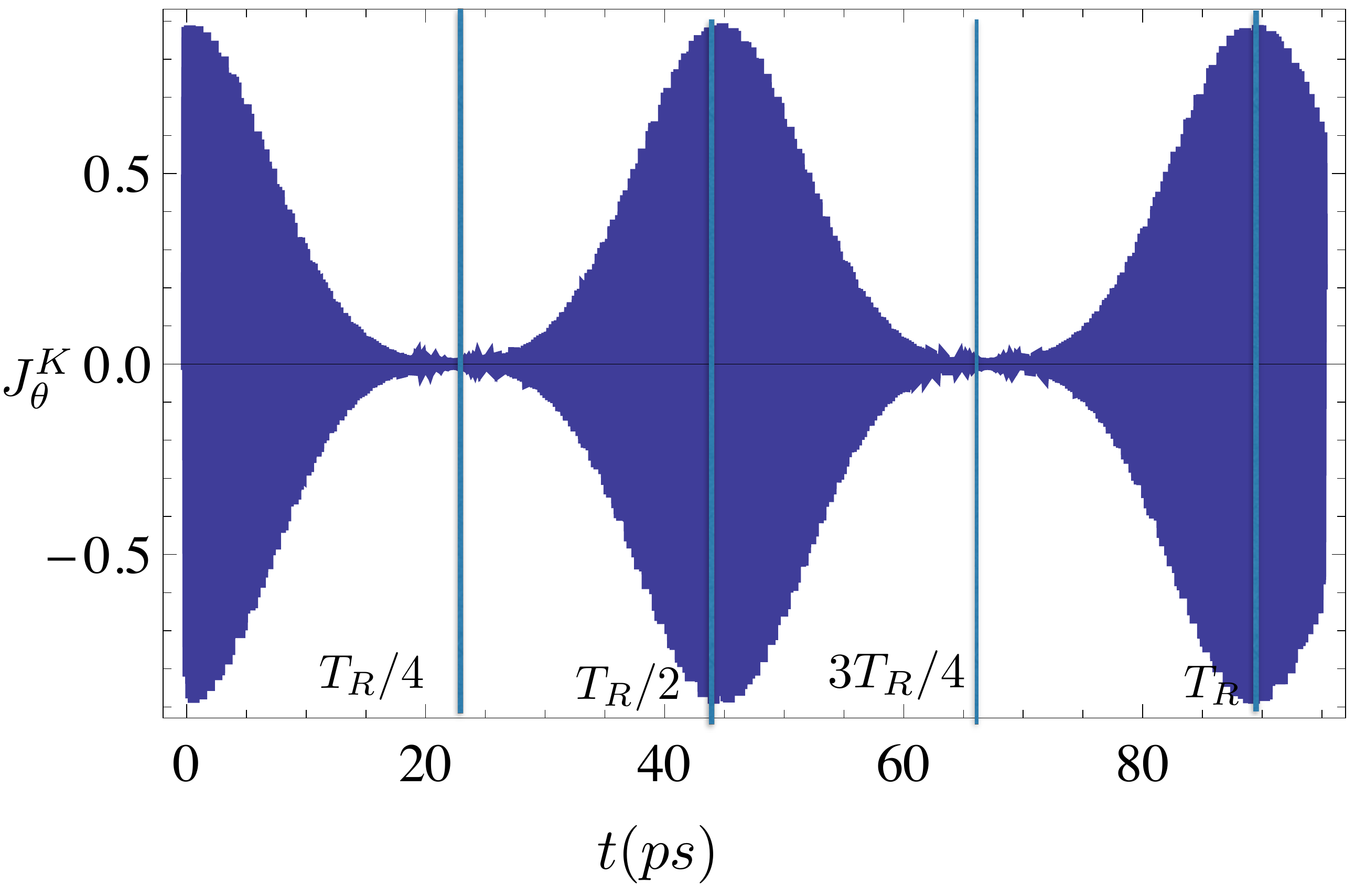}}
\caption{Time evolution of $J^{K}_{\theta}$ for a pentagonal defect (upper pannel) and a square defect (lower pannel). The difference lies in the time scales (horizontal axis).}
\label{sq_current_osc}
\end{figure}
\begin{figure}
\rotatebox{0}{\includegraphics[width=3.2in]{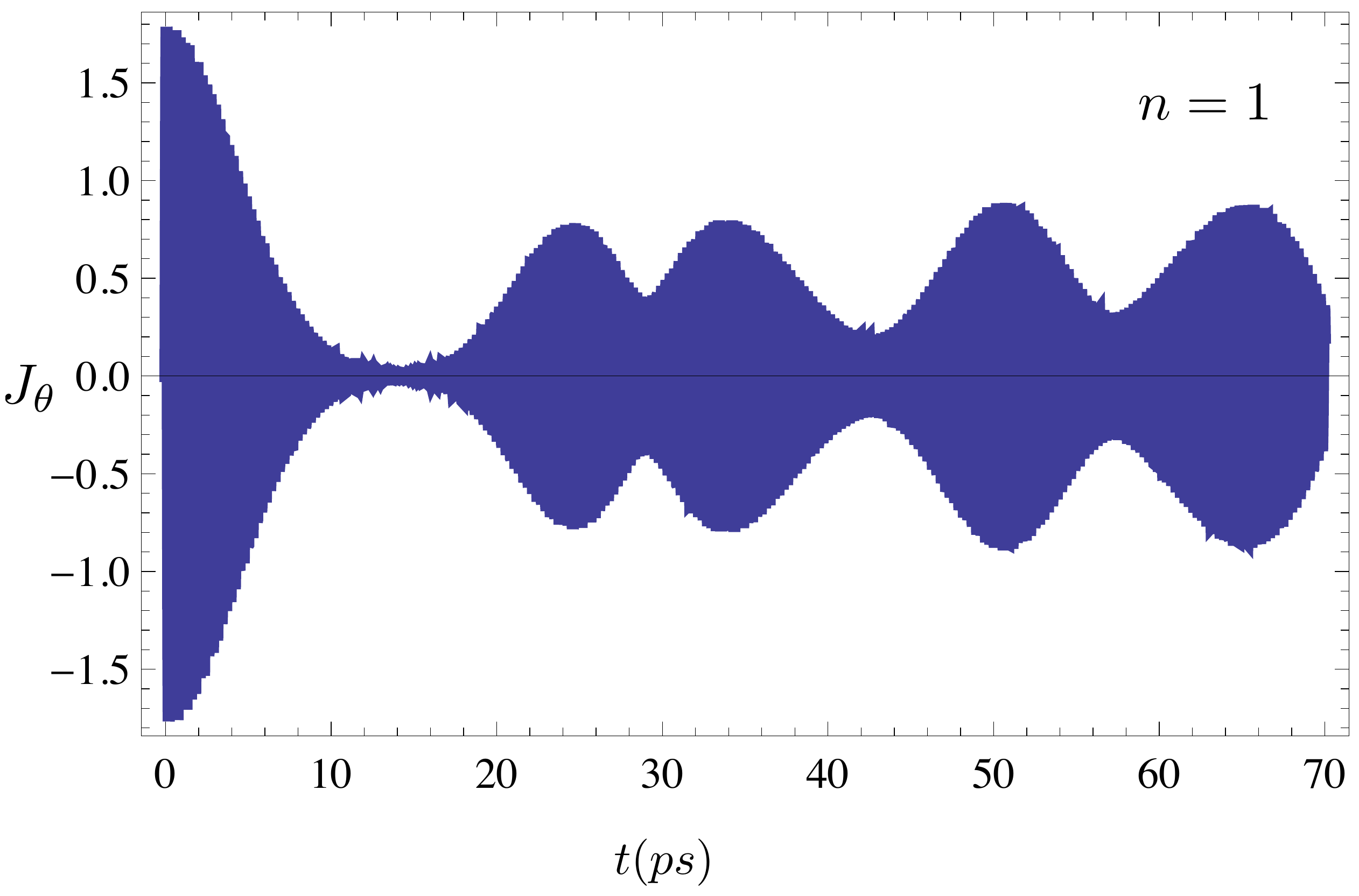}}
\caption{Time evolution of total angular current $J_{\theta}$ for a pentagon defect.}
\label{total_pent_current_osc}
\end{figure}

It is also seen from the results listed in table \ref{table} that the classical and revival times are different at $K$ and $K'$. This difference in time scales actually destroys the oscillating behaviour of the total charge current as shown in Fig.~\ref{total_pent_current_osc} for a pentagonal defect in graphene. This unusual behaviour is one of the main results of this paper.  

%

To investigate the time evolution of angular current $J_{\theta}$ in Eq.~(\ref{total_current}), we expand in the base of $\phi_{m}(\rho,\theta)$. One finds for the expected temporal evolution of the current \cite{Romera2}
\begin{eqnarray}
J^{K}_{\theta}=v_{F}\sum^{\infty}_{m=1}|c_{m,m-1}|^2\sin[(E_{m}-E_{m-1})t/\hbar]
\end{eqnarray}
One can find $E_{m}-E_{m-1}\simeq E'_{m_{0}}+E''_{m_{0}}(m-m_{0})$ from which the scales $T_{Cl}$ and $T_{R}$ are arise. It is shown in Fig. 
(\ref{sq_current_osc}) the time behaviour of the current $J_{\theta}$ for defects of indices $n=1$ and $n=2$. When after a few periods the initial wave packet enters the collapse phase and the quasiclassical oscillatory behaviour of the currents vanishes, only to emerge later at half the revival time.


\section{Summary and Conclusions}\label{Summary}

In this paper, we have shown how a specific type of  topological defect, a disclination which produces an effective gauge field coupled to the graphene electrons can modify the physical properties of the material by the breaking of valley degeneracies. 

\begin{itemize}
\item The presence of the disclination (the effect of which is measured by a non-vanishing integer index $n$) modifies the energy spectrum (which in particular can be gapless at one Dirac point and gapped at the other) and has the effect of shifting the magnetic flux in opposite directions at the two Dirac points $K$ and $K'$. 

\item As a consequence, there is a modification of the charge current, and the fictitious magnetic field produced by the disclination having opposite signs at the two Dirac points, the currents there partially cancel each other. The total current as a half periodicity. Although for even $n$ this scenario is different, contribution from two Dirac points are in same phase and enhance the total charge currents. 

\item The disclination may also enhance or smoothen (depending on the sign of $n$) the pseudo-spin polarization which is a contribution to the orbital angular momentum.

\item This also leads to remarkable  features on quantum oscillations and revival times of wave packets, for example the fact that the revival times are not exactly identical at the two Dirac points. As a consequence, the oscillating behavior of the current is partially destroyed with time. This is true for any type disclinations in graphene.
\end{itemize}

There should exist experimental signatures of these features. For example in the presence of a single square disclination, the graphene sheet  is gapped at $K-$point but gapless at $K'$ and thus the conductivity of the metallic state should be half that of the pure graphene sheet. The change of periodicity in the current dependence on the magnetic flux is even a more obvious signature.

\section{Acknowledgement} We thank Ernesto Medina for useful correspondence. D.S. like to thank the Statistical Physics Group in the University the Lorraine  where part of this work has been done for  hospitality.

\end{document}